\documentclass[aps,pre,twocolumn,groupedaddress,showpacs]{revtex4-1}

\usepackage[table]{xcolor}
\usepackage{graphicx}

\begin{document}


\title{Multivariate information measures: an experimentalist's perspective}


\author{Nicholas Timme}
\email{nmtimme@umail.iu.edu}
\affiliation{Department of Physics, Indiana University, Bloomington, IN 47405-7105}
\author{Wesley Alford}
\affiliation{Department of Physics, Indiana University, Bloomington, IN 47405-7105}
\author{Benjamin Flecker}
\affiliation{Department of Physics, Indiana University, Bloomington, IN 47405-7105}
\author{John M. Beggs}
\affiliation{Department of Physics, Indiana University, Bloomington, IN 47405-7105}


\date{\today}

\begin{abstract}
Information theory has long been used to quantify interactions between two variables. With the rise of complex systems research, multivariate information measures are increasingly needed. Although the bivariate information measures developed by Shannon are commonly agreed upon, the multivariate information measures in use today have been developed by many different groups, and differ in subtle, yet significant ways. Here, we will review the information theory behind each measure, as well as examine the differences between these measures by applying them to several simple model systems. In addition to these systems, we will illustrate the usefulness of the information measures by analyzing neural spiking data from a dissociated culture through early stages of its development. We hope that this work will aid other researchers as they seek the best multivariate information measure for their specific research goals and system. Finally, we have made software available online which allows the user to calculate all of the information measures discussed within this paper.
\end{abstract}


\maketitle

\section{Introduction}

Information theory has proved to be a useful tool in many disciplines. It has been successfully applied in several areas of research, including neuroscience \cite{Rieke1997}, data compression \cite{Ziv1977}, coding \cite{Berrou1993}, dynamical systems \cite{Fraser1986}, and genetic coding \cite{Butte2000}, just to name a few. Information theory's broad applicability is due in part to the fact that it relies only on the probability distribution associated with one or more variables. Generally speaking, information theory uses the probability distributions associated with the values of the variables to ascertain whether or not the values of the variables are related and, depending on the situation, the way in which they are related. As a result of this, information theory can be applied to linear and non-linear systems, although this does not guarantee that an information-based measure will capture all nonlinear contributions. In summary, information theory is a model-independent approach.

Information theoretic approaches to problems involving one and two variables are well understood and widely used. In addition to the one and two variable measures, several information measures have been introduced to analyze the relationships or interactions between three or more variables \cite{McGill1954,Watanabe1960,Han1975,Chechik2001,Nirenberg2001,SchneidmanStill2003,Varadan2006,Williams2010}. These multivariate information measures have been applied in physical systems \cite{Cerf1997,Matsuda2000}, biological systems \cite{Anastassiou2007,Chanda2007}, and neuroscience \cite{Brenner2000,SchneidmanBialek2003,Bettencourt2007}. However, these multivariate information measures differ in significant and sometimes subtle ways. Furthermore, the notation and naming associated with these measures is inconsistent throughout the literature (see, for example, \cite{Watanabe1960,Tononi1994,Sporns2000,SchneidmanStill2003,Wennekers2003}). Within this paper, we will examine a wide array of multivariate information measures in an attempt to clearly articulate the different measures and their uses. After reviewing the information theory behind each individual measure, we will apply the information measures to several model systems in order to illuminate their differences and similarities. Also, we will apply the information measures to neural spiking data from a dissociated neural culture. Our goal is to clarify these methods for other researchers as they search for the multivariate information measure that will best address their specific research goals.

In order to facilitate the use of the information measures discussed in this paper, we have made our MATLAB software freely available, which can be used to calculate all of the information measures discussed herein \cite{TimmeWebsite}. An earlier version of this work was previously posted on the arXiv \cite{Timme2011}.

\section{Synergy and redundancy}

A crucial topic related to multivariate information measures is the distinction between synergy and redundancy. With regard to these information measures, the precise meanings of ``synergy'' and ``redundancy'' have not been established, though they have been invoked by many researchers in this field (see, for instance, \cite{Brenner2000,Williams2010}). For a recent treatment of synergy in this context, see \cite{Griffith2011}. 

To begin to understand synergy, we can use a simple system. Suppose two variables (call them $X_1$ and $X_2$) provide some information about a third variable (call it $Y$). In other words, if you know the state of $X_1$ and $X_2$, then you know something about the state $Y$. Loosely, the portion of that information that is not provided by knowing both $ X_1$ alone and $X_2$ alone is said to be provided synergistically by $X_1$ and $X_2$. The synergy is the bonus information received by knowing $X_1$ and $X_2$ \emph{together}, instead of separately. 

We can take a similar initial approach to redundancy. Again, suppose $X_1$ and $X_2$ provide some information about $Y$. The common portion of the information $X_1$ provides alone and the information $X_2$ provides alone is said to be provided redundantly by $X_1$ and $X_2$. The redundancy is the information received from \emph{both} $X_1$ and $X_2$. 

These imprecise definitions may seem clear enough, but in attempting to measure these quantities, researchers have created distinct measures that produce different results. Based on the fact that the overall goal has not been clearly defined, it cannot be said that one of these measures is ``correct.'' Rather, each measure has its own uses and limitations. Using the simple systems below, we will attempt to clearly articulate the differences between the multivariate information measures.

\section{Multivariate information measures}

In this section we will discuss the various multivariate information theoretic measures that have been introduced previously. Of special note is the fact that the names and notation used in the literature have not been consistent. We will attempt to clarify the discussion as much as possible by listing alternative names when appropriate. We will refer to an information measure by its original name (or at least, its original name to the best of our knowledge).

\subsection{Entropy and mutual information}

The information theoretic quantities involving one and two variables are well-defined and their results are well-understood. Regarding the probability distribution of one variable (call it $p(x)$), the canonical measure is the entropy $H(x)$ \cite{Cover2006}. The entropy is given by \footnote{Throughout the paper we will use capital letters to refer to variables and lower case letters to refer to individual values of those variables. We will also use discrete variables, though several of the information measures discussed can be directly extended to continuous variables. When working with a continuous variable, various techniques exists, such as kernel density estimation, which can be used to infer a discrete distribution from a continuous variable. Logarithms will be base 2 throughout in order to produce information values in units of bits.}: 
\begin{equation}\label{EQ1}
H(X) \equiv -\sum_{x \in X}p(x)\log(p(x))
\end{equation}
The entropy quantifies the amount of uncertainty that is present in the probability distribution. If the probability distribution is concentrated near one value, the entropy will be low. If the probability distribution is uniform, the entropy will be at a maximum. 

When examining the relationship between two variables, the mutual information (I) quantifies the amount of information provided about one of the variables by knowing the value of the other \cite{Cover2006}. The mutual information is given by:
\begin{eqnarray}\label{EQ2}
I(X;Y) \equiv H(X)-H(X|Y)=H(Y)-H(Y|X)= \nonumber \\
H(X)+H(Y)-H(X,Y) \hspace{1cm}
\end{eqnarray}
where the conditional entropy is given by:
\begin{eqnarray}\label{EQ3}
H(X|Y)=\sum_{y \in Y}p(y)H(X|y) \hspace{1cm} \nonumber \\
=\sum_{y \in Y}p(y)\sum_{x \in X}p(x|y)\log\frac{1}{p(x|y)} \hspace{0.5cm}
\end{eqnarray}

The mutual information can also be written as the Kullback-Leibler divergence between the joint probability distribution of the actual data and the joint probability distribution of the independent model (wherein the joint distribution is equal to the product of the marginal distributions). This form is given by:
\begin{equation}\label{EQ4}
I(X;Y)=\sum_{x \in X,y \in Y}p(x,y)\log\left(\frac{p(x,y)}{p(x)p(y)}\right)
\end{equation}

The mutual information can be used as a measure of the interactions among more than two variables by grouping the variables into sets and treating each set as a single vector-valued variable. For instance, the mutual information can be calculated between $Y$ and the set $S = \{X_1,X_2\}$\footnote{We will use S to refer to a set of $N$ $X$ variables such that $S = \{X_1, X_2, \ldots X_N\}$ throughout the paper.}  in the following way:
\begin{equation}\label{EQ5}
I(Y;S)=\sum_{y \in Y \atop x_1 \in X_1,x_2 \in X_2}p(y,x_1,x_2)\log\left(\frac{p(y,x_1,x_2)}{p(y)p(x_1,x_2)}\right)
\end{equation}
However, when the mutual information is considered as in Eq. (\ref{EQ5}), it is not possible to separate contributions from individual $X$ variables in the set $S$. Still, by varying the number of variables in $S$, the mutual information in Eq. (\ref{EQ5}) can be used to measure the gain or loss in information about $Y$ by those variables in $S$. Along these lines, Bettencourt et al. used the mutual information between one variable (in their case, the activity of a neuron) and many other variables considered together (in their case, the activities of a group of other neurons) in order to examine the relationship between the amount of information the group of neurons provided about the single neuron to the number of neurons considered in the group \cite{Bettencourt2008}. 

The mutual information can be conditioned upon a third variable to yield the conditional mutual information \cite{Cover2006}. It is given by:
\begin{eqnarray}\label{EQ6}
I(X;Y|Z)= \hspace{4cm} \nonumber \\
\sum_{z \in Z}p(z)\sum_{x \in X,y \in Y}p(x,y|z)\log\frac{p(x,y|z)}{p(x|z)p(y|z)}= \hspace{0.25cm} \nonumber \\
\sum_{x \in X,y \in Y,z \in Z}p(x,y,z)\log\frac{p(z)p(x,y,z)}{p(x,z)p(y,z)}
\end{eqnarray}
The conditional mutual information quantifies the amount of information one variable provides about a second variable when a third variable is known. 

\subsection{Interaction information}

The first attempt to quantify the relationship among three variables in a joint probability distribution was the interaction information (II), which was introduced by McGill \cite{McGill1954}. It attempts to extend the concept of the mutual information as the information gained about one variable by knowing the other. The interaction information is given by:
\begin{eqnarray}\label{EQ7}
II(X;Y;Z)\equiv I(X,Y|Z)-I(X;Y)= \hspace{1.5cm} \nonumber \\
I(X,Z|Y)-I(X;Z)=I(Z,Y|X)-I(Z,Y) \hspace{0.25cm}
\end{eqnarray}

Of the interaction information, McGill said \cite{McGill1954}, ``We see that $II(X;Y;Z)$ is the gain (or loss) in sample information transmitted between any two of the variables, due to the additional knowledge of the third variable.'' The interaction information can also be written as:
\begin{equation}\label{EQ8}
II(X;Y;Z)=I(X,Y;Z)-(I(X;Z)+I(Y;Z))
\end{equation}

In the form given in Eq. (\ref{EQ8}), the interaction information has been widely used in the literature and has often been referred to as the synergy \cite{Gat1999,Brenner2000,SchneidmanBialek2003,Anastassiou2007} and redundancy-synergy index \cite{Chechik2001}. Some authors have used the term ``synergy'' because they have interpreted a positive interaction information result to imply a synergistic interaction among the variables and a negative interaction information result to imply a redundant interaction among the variables. Thus, if we assume this interpretation of the interaction information and that the interaction information correctly measures multivariate interactions, then synergy and redundancy are taken to be mutually exclusive qualities of the interactions between variables. This view will find a counterpoint in the partial information decomposition to be discussed below. 

The interaction information can also be written as an expansion of the entropies and joint entropies of the variables:
\begin{eqnarray}\label{EQ9}
II(X;Y;Z)=-H(X)-H(Y)-H(Z)+ \hspace{1.25cm} \nonumber \\
H(X,Y)+H(X,Z)+H(Y,Z)-H(X,Y,Z) \hspace{0.5cm}
\end{eqnarray}

This form leads to an expansion for the interaction information for $N$ number of variables \cite{Jakulin2008}. If $S = \{X_1, X_2, \ldots X_N\}$,  then the interaction information becomes:
\begin{equation}\label{EQ10}
II(S)=-\sum_{T \subseteq S}(-1)^{|S|-|T|}H(T)
\end{equation}
In Eq. (\ref{EQ10}), $T$ is a subset of $S$ and $|S|$ denotes the set size of $S$.

A measure similar to the interaction information was introduced by Bell and is referred to as the co-information (CI) \cite{Bell2003}. It is given by the following expansion:
\begin{equation}\label{EQ11}
CI(S) \equiv -\sum_{T \subseteq S}(-1)^{|T|}H(T)=(-1)^{|S|}II(S)
\end{equation}

Clearly, the co-information is equal to the interaction information when $S$ contains an even number of variables and is equal to the negative of the interaction information when $S$ contains an odd number of variables. So, for the three variable case, the co-information becomes:
\begin{eqnarray}\label{EQ12}
CI(X;Y;Z)=I(X;Y)-I(X,Y|Z)= \hspace{0.75cm} \nonumber \\
I(X;Z)+I(Y;Z)-I(X,Y;Z) \hspace{0.5cm}
\end{eqnarray}

Because the co-information is directly related to the interaction information for systems with any number of variables, we will forgo presenting results from the co-information. The co-information has also been referred to as the generalized mutual information \cite{Matsuda2000}.

\subsection{Total correlation}

The interaction information finds its conceptual base in extending the idea of the mutual information as the information gained about a variable when the other variable is known. Alternatively, we could extend the idea of the mutual information as the Kullback-Leibler divergence between the joint distribution and the independent model. If we do this, we arrive at the total correlation (TC) introduced by Watanabe \cite{Watanabe1960}. It is given by:
\begin{equation}\label{EQ13}
TC(S) \equiv \sum_{\vec{x} \in S}p(\vec{x})\log\left(\frac{p(\vec{x})}{p(x_1)p(x_2) \ldots p(x_n)}\right)
\end{equation}
In Eq. (\ref{EQ13}), $\vec{x}$ is a vector containing individual states of the $X$ variables. The total correlation can also be written in terms of entropies as:
\begin{equation}\label{EQ14}
TC(S) =\left(\sum_{X_i \in S}H(X_i)\right)-H(S)
\end{equation}
In this form, the total correlation has been referred to as the multi-information \cite{SchneidmanStill2003}, the spatial stochastic interaction \cite{Wennekers2003}, and the integration \cite{Tononi1994,Sporns2000}. Using Eq. (\ref{EQ2}), the total correlation can also be written using a series of mutual information terms (see Appendix \ref{Ap1} for more details):
\begin{eqnarray}\label{EQ29}
TC(S)=I(X_1;X_2)+I(X_1,X_2;X_3)+ \ldots \nonumber \\
I(X_1, \ldots ,X_{n-1};X_n) \hspace{2.5cm}
\end{eqnarray}

\subsection{Dual total correlation}

After the total correlation was introduced, a measure with a similar structure, called the dual total correlation (DTC), was introduced by Han \cite{Han1975,Han1978}. The dual total correlation is given by:
\begin{equation}\label{EQ30}
DTC(S) \equiv \left(\sum_{X_i \in S}H(S/X_i)\right)-(n-1)H(S)
\end{equation}

In Eq. (\ref{EQ30}), $S/X_i$ is the set $S$ with $X_i$ removed and $n$ is the number of $X$ variables in $S$. The dual total correlation can also be written as \cite{Abdallah2010}:
\begin{equation}\label{EQ31}
DTC(S)=H(S)-\sum_{X_i \in S}H(X_i|S/X_i)
\end{equation}

The dual total correlation calculates the amount of entropy present in $S$ beyond the sum of the entropies for each variable conditioned upon all other variables. The dual total correlation has also been referred to as the excess entropy \cite{Olbrich2008} and the binding information \cite{Abdallah2010}. Using Eq. (\ref{EQ2}), (\ref{EQ14}), and (\ref{EQ30}), the dual total correlation can also be related to the total correlation by (see Appendix \ref{Ap2} for more details):
\begin{equation}\label{EQ33}
DTC(S)=\left(\sum_{X_i \in S}I(S/X_i;X_i)\right)-TC(S)
\end{equation}

\subsection{$\Delta I$}

A distinct information measure, called $\Delta I$, was introduced by Nirenberg and Latham \cite{Nirenberg2001,Latham2005}. It was introduced to measure the importance of correlations in neural coding. For the purposes of this paper, we can apply $\Delta I$ to the following situation: consider some set of $X$ variables (call this set $S$). The values of the variables in $S$ are related in some way to the value of another variable (call it $Y$). In Nirenberg and Latham's original work, the $X$ variables are signals from neurons and the $Y$ variable is the value of some stimulus variable. $\Delta I$ compares the true probability distributions associated with these variables to one that assumes the $X$ variables act independently (i.e., there are no correlations between the $X$ variables beyond those that can be explained by $Y$). If these distributions are similar, then it can be assumed that there are no relevant correlations between the $X$ variables. If, on the other hand, these distributions are not similar, then we can conclude that relevant correlations are present between the $X$ variables. 

The independent model assumes that the $X$ variables act independently, so we can form the probability for the $X$ states conditioned upon the $Y$ variable state using a simple product:
\begin{equation}\label{EQ15}
p_{ind}(\vec{x}|y)=\prod_ip(x_i|y)
\end{equation}
Then, the conditional probability of the $Y$ variable on the $X$ variables can be found using Bayes' theorem.
\begin{equation}\label{EQ16}
p_{ind}(y|\vec{x})=\frac{p_{ind}(\vec{x}|y)p(y)}{p_{ind}(\vec{x})}
\end{equation}

The independent joint distribution of the $X$ variables is given by:
\begin{equation}\label{EQ17}
p_{ind}(\vec{x})=\sum_{y \in Y}p_{ind}(\vec{x}|y)p(y)
\end{equation}
Then, $\Delta I$ is given by the weighted Kullback-Leibler distance between the conditional probability of the $Y$ variable on the $X$ variables for the independent model and the actual conditional probability of the same type.
\begin{equation}\label{EQ18}
\Delta I(S;Y) \equiv \sum_{\vec{x} \in S}p(\vec{x})\sum_{y \in Y}p(y|\vec{x})\log\left(\frac{p(y|\vec{x})}{p_{ind}(y|\vec{x})}\right)
\end{equation}

About $\Delta I$, Nirenberg and Latham say \cite{Latham2005}, ``[s]pecifically, $\Delta I$ is the cost in yes/no questions for not knowing about correlations: if one were guessing the value of the $Y$ variable based on the $X$ variables, $\vec{x}$, then it would take, on average, $\Delta I$ more questions to guess the value of $Y$ if one knew nothing about the correlations than if one knew everything about them [Variable names changed to match this work].''

\subsection{Redundancy-synergy index}

Another multivariate information measure was introduced by Chechik et al. \cite{Chechik2001}. This measure was originally referred to as the redundancy-synergy index (RSI) and it was created as an extension of the interaction information. It is given by:
\begin{equation}\label{EQ19}
RSI(S;Y) \equiv I(S;Y)-\sum_{X_i \in S}I(X_i;Y)
\end{equation}

The redundancy-synergy index is designed to be maximal and positive when the variables in $S$ are purported to provide synergistic information about $Y$. It should be negative when the variables in $S$ provide redundant information about $Y$. When $S$ contains two variables, the redundancy-synergy index is equal to the interaction information. The negative of the redundancy-synergy index has also been referred to as the redundancy \cite{SchneidmanStill2003}.

\subsection{Varadan's synergy}

Yet another multivariate information measure was introduced by Varadan et al. \cite{Varadan2006}. In the original work, this measure is referred to as the synergy, but to avoid confusing it with other measures, we will refer to this measure as Varadan's synergy (VS). It is given by:
\begin{equation}\label{EQ20}
VS(S;Y) \equiv I(S;Y)-max\sum_jI(S_j;Y)
\end{equation}
In Eq. (\ref{EQ20}), $S_j$ refers to the possible sub-sets of $S$. So, for instance, if $S = \{X_1, X_2, X_3\}$, Varadan's synergy would be given by:
\begin{eqnarray}\label{EQ21}
VS(S;Y)=I(S;Y) \hspace{3.5cm} \nonumber \\
-max\left\{
\begin{array}{c}
I(X_1;Y)+I(X_2,X_3;Y) \\
I(X_1;Y)+I(X_2,X_3;Y) \\
I(X_1;Y)+I(X_2,X_3;Y) \\
I(X_1;Y)+I(X_2;Y)+I(X_3;Y)
\end{array} \right.
\end{eqnarray}

Similar to the interaction information, when Varadan's synergy is positive, the variables in $S$ are said to provide synergistic information about $Y$, while when Varadan's synergy is negative, the variables in $S$ are said to provide redundant information about $Y$. Note that, when $S = \{X_1, X_2\}$, Varadan's synergy is equal to the interaction information. 

\subsection{Partial information decomposition}

Finally, we will examine the collection of information values introduced by Williams and Beer in the partial information decomposition (PID) \cite{Williams2010}.  (For three other applications of the partial information decomposition, see recent works by James et al. \cite{James2011}, Flecker et al. \cite{Flecker2011}, and Griffith and Koch \cite{Griffith2011}). The partial information decomposition is a method of dissecting the mutual information between a set of variables $S$ and one other variable $Y$ into non-overlapping terms. These terms quantify the information provided by the set of variables in $S$ about $Y$ uniquely, redundantly, synergistically, and in mixed forms. The partial information decomposition has several potential advantages over other measures. First, it produces only non-negative results, unlike the interaction information. Second, it allows for the possibility of synergistic and redundant interactions simultaneously, unlike the interaction information and $\Delta I$.

For the sake of brevity, we will not describe the entire partial information decomposition here, but we will describe the case where $S = \{X_1, X_2\}$. A description of the general case can be found in Williams and Beer's original work \cite{Williams2010}. The relevant mutual informations are equal to sums of the partial information terms. For the case of two $X$ variables, there are only four possible terms. Information about $Y$ can be provided uniquely by each $X$ variable, redundantly by both $X$ variables, or synergistically by both $X$ variables together. Written out, the relevant mutual informations are given by the following sums:
\begin{eqnarray}\label{EQ22}
I(X_1,X_2;Y)=Synergy(X_1,X_2)+Unique(X_1)+ \nonumber \\
Unique(X_2)+Redundancy(X_1,X_2) \hspace{1.5cm}
\end{eqnarray}

\begin{equation}\label{EQ23}
I(X_1;Y)=Unique(X_1)+Redundancy(X_1,X_2)
\end{equation}

\begin{equation}\label{EQ24}
I(X_2;Y)=Unique(X_2)+Redundancy(X_1,X_2)
\end{equation}

The relevant mutual information values can be calculated easily. As described by Williams and Beer, the redundancy term is equal to a new information expression: the minimum information function. This function attempts to capture the intuitive view that the redundant information for a given state of $Y$ is the information that is contributed by both $X$ variables about that state of $Y$ (consult Williams and Beer's original work \cite{Williams2010} for details and further motivation). The minimum information function is related to the specific information \cite{DeWeese1999}\footnote{It should be noted that DeWeese and Meister refer to the expression in Eq. (\ref{EQ25}) as the specific surprise.}. The specific information is given by:
\begin{equation} \label{EQ25}
I_{spec}(y;X)=\sum_{x \in X} p(x \mid y)\left[ log\left(\frac{1}{p(y)}\right) - log \left(\frac{1}{p(y \mid x)}\right) \right]
\end{equation}
In Eq. (\ref{EQ25}), the specific information quantifies the amount of information provided by X about a specific state of the Y variable. 

The minimum information can then be calculated by comparing the amount of information provided by the different $X$ variables for each state of the $Y$ variable considered individually.
\begin{equation} \label{EQ26}
I_{min}(Y;X_1,X_2)=\sum_{y \in Y}p(y)min_{X_i}I_{spec}(y;X_i)
\end{equation}
The minimum in Eq. (\ref{EQ26}) is taken over each $X$ variable considered separately. Once the redundancy term is calculated via the minimum information function, the remaining partial information terms can be calculated with ease. 

It should also be noted that the partial information decomposition provides an explanation for negative interaction information values. To see this, insert the partial information expansions in Eq. (\ref{EQ22}), (\ref{EQ23}), and (\ref{EQ24}) to the mutual information terms in the interaction information:
\begin{eqnarray} \label{EQ27}
II(X_1;X_2;Y)=I(X_1,X_2;Y)-I(X_1,Y)-I(X_2,Y)= \nonumber \\
Synergy(X_1,X_2) - Redundancy(X_1,X_2) \hspace{1.5 cm}
\end{eqnarray}
Thus, the partial information decomposition finds that a negative interaction information value implies that the redundant contribution is greater than the synergistic contribution. Furthermore, the structure of the partial information decomposition implies that synergistic and redundant interactions are not mutually exclusive, as was the case for the traditional interpretation of the interaction information. Thus, according to the partial information decomposition, there may be non-zero synergistic and redundant contributions simultaneously. 

Throughout the remainder of this article, we will label the various terms in the partial information decomposition in accordance with the notation used by Williams and Beer. The term that has been interpreted as the synergy will be referred to as $\Pi_R(Y;\{12\})$ or PID synergy. The term that has been interpreted as the redundancy will be labeled as $\Pi_R(Y;\{1\}\{2\})$ or PID redundancy. The unique information terms will be referred to as $\Pi_R(Y;\{1\})$ and $\Pi_R(Y;\{2\})$, or simply as PID unique information. 

When the partial information decomposition is extended to the case where $S = \{X_1, X_2, X_3\}$, new mixed terms are introduced to the expansions of the mutual informations. For instance, information can be supplied about $Y$ redundantly between $X_3$ and the synergistic contribution from $X_1$ and $X_2$ (this term is noted as $\Pi_R(Y;\{12\}\{3\})$). In total, the partial information decomposition contains 18 terms when $S$ contains three variables. It can be shown that the interaction information between $Y$ and the $X$ variables contained in $S$ is related to the partial information terms by the following equation \cite{Williams2010}:
\begin{eqnarray} \label{EQ28}
II(Y;X_1;X_2;X_3)= \Pi_R(Y;\{123\})+\hspace{1cm}\nonumber \\
\Pi_R(Y;\{1\}\{2\}\{3\})-\Pi_R(Y;\{1\}\{23\})- \nonumber \\
\Pi_R(Y;\{2\}\{13\})-\Pi_R(Y;\{3\}\{12\})- \nonumber \\
\Pi_R(Y;\{12\}\{13\})-\Pi_R(Y;\{12\}\{23\})- \nonumber \\
\Pi_R(Y;\{13\}\{23\})-2\Pi_R(Y;\{12\}\{13\}\{23\})
\end{eqnarray}
From Eq. (\ref{EQ28}), we can see that the four-way interaction information is related to the partial information decomposition via a complicated summation of terms. 

\section{Example systems}

We will now apply the multivariate information measures discussed above to several simple systems in an attempt to understand their similarities, differences, and uses. These systems have been chosen to maximize the contrast between the information measures, but many other systems exist for which the information measures produce identical results. 

\subsection{Examples 1-3: two-input Boolean logic gates}

The first set of examples we will consider are simple Boolean logic gates. These logic gates are well known across many disciplines and offer a great deal of simplicity. The results presented in Table \ref{TableI} highlight some of the commonalities and disparities between the various information measures. It should be noted that, due to the simple structure of the Boolean logic gates, the total correlation is equal to the mutual information. Also, due to the fact that only two-input Boolean logic gates are being considered, the redundancy-synergy index and Varadan's synergy are directly related to the interaction information. Additional examples will highlight differences between these information measures.

\begin{table}
\caption{Examples 1 to 3: two-input Boolean logic gates. XOR-gate: All information measures produce consistent results. $X_1$-gate: The partial information decomposition succinctly identifies a relationship between $X_1$ and $Y$. AND-gate: The partial information decomposition identifies both synergistic and redundant interactions. The interaction information finds only a synergistic interaction. $\Delta I$ identifies the importance of correlations between $X_1$ and $X_2$. \label{TableI}}
\begin{tabular}{>{\hfill}ccc|c|c|c} 
\multicolumn{3}{c|}{ } & XOR & $X_1$ & AND \\ \hline
$p(x_1,x_2,y)$ & $x_1$ & $x_2$ & $y$ & $y$ & $y$ \\ \hline
$1/4$ & $0$ & $0$ & $0$ & $0$ & $0$ \\ 
$1/4$ & $1$ & $0$ & $1$ & $1$ & $0$ \\ 
$1/4$ & $0$ & $1$ & $1$ & $0$ & $0$ \\ 
$1/4$ & $1$ & $1$ & $0$ & $1$ & $1$ \\ \hline 
\multicolumn{3}{c|}{$I(X_1;Y)$} & $0$ & $1$ & $0.311$ \\ 
\multicolumn{3}{c|}{$I(X_2;Y)$} & $0$ & $0$ & $0.311$ \\ 
\multicolumn{3}{c|}{$I(X_1,X_2;Y)$} & $1$ & $1$ & $0.811$ \\ 
\multicolumn{3}{c|}{$II(X_1;X_2;Y)$} & $1$ & $0$ & $0.189$ \\ 
\multicolumn{3}{c|}{$TC(X_1;X_2;Y)$} & $1$ & $1$ & $0.811$ \\ 
\multicolumn{3}{c|}{$DTC(X_1;X_2;Y)$} & $2$ & $1$ & $1$ \\ 
\multicolumn{3}{c|}{$\Delta I(X_1,X_2;Y)$} & $1$ & $0$ & $0.104$ \\ 
\multicolumn{3}{c|}{$RSI(X_1,X_2;Y)$} & $1$ & $0$ & $0.189$ \\ 
\multicolumn{3}{c|}{$VS(X_1,X_2;Y)$} & $1$ & $0$ & $0.189$ \\ 
\multicolumn{3}{c|}{$\Pi_R(Y;\{1\}\{2\})$} & $0$ & $0$ & $0.311$ \\ 
\multicolumn{3}{c|}{$\Pi_R(Y;\{1\})$} & $0$ & $1$ & $0$ \\ 
\multicolumn{3}{c|}{$\Pi_R(Y;\{2\})$} & $0$ & $0$ & $0$ \\ 
\multicolumn{3}{c|}{$\Pi_R(Y;\{12\})$} & $1$ & $0$ & $0.5$ \\ 
\end{tabular}
\end{table}

All information measures provide a similar result for the XOR-gate (with the exception of the dual total correlation, see below). The interaction information, $\Delta I$, the redundancy-synergy index, Varadan's synergy, and the partial information decomposition all indicate that the entire bit of information between $Y$ and $\{X_1,X_2\}$ is accounted for by synergy. We might expect this result because, to know the state of $Y$ for an XOR-gate, the state of both $X_1$ and $X_2$ must be known. 

The results for $X_1$ gate demonstrate the potential utility of the partial information decomposition. The unique information term from $X_1$ is equal to one bit, thus indicating that the $X_1$ variable entirely and solely determines the state of the output variable. This result is confirmed by the truth-table. This result can also be seen by considering the values of the other measures together (for instance, the three mutual information measures), but the partial information decomposition provides these results more succinctly. 

More significant differences among the information measures appear when considering the AND-gate. The partial information decomposition produces the result that 0.311 bits of information are provided redundantly and 0.5 bits are provided synergistically. Since each $X$ variable provides the same amount of information about each state of $Y$ (see Eq. (\ref{EQ26})), the partial information decomposition finds that all of the mutual information between each $X$ variable individually and the $Y$ variable is redundant. As a result of this, no information is provided uniquely, and subsequently, the entirety of the remaining 0.5 bits of information between $Y$ and $\{X_1,X_2\}$ must be synergistic. From this, we can see in action the fact that the partial information decomposition emphasizes the amount of information that each $X$ variable provides about each state of $Y$ \emph{considered individually}. 

The interaction information, and by extension the redundancy-synergy index and Varadan's synergy, are limited to returning only a synergy value of 0.189 bits for the AND-gate. This value is produced because the mutual information between $Y$ and $\{X_1,X_2\}$ contains an excess of 0.189 bits beyond the sum of the mutual informations between each $X$ variable individually and the $Y$ variable. So, here we can see in action the interpretation of the interaction information as the amount of information provided by the $X$ variables taken together about $Y$, beyond what they provide individually. Also, the AND-gate allows us to see the relationship between the interaction information and the partial information decomposition as expressed by Eq. (\ref{EQ27}).

The value of $\Delta I$ for the AND-gate can be elucidated by examining the values of the conditional probability distributions that are relevant to the calculation of $\Delta I$ (Table \ref{TableII}). From these results, it is clear that if we use the independent model, and we are presented with the state $x_1 = 1$ and $x_2 = 1$, we would conclude that there is a one-quarter chance that $y = 0$ and a three-quarters chance that $y = 1$. If we use the actual data, then we know that, for that specific state, $y$ must equal $1$. This example points to a subtle, but critical difference between $\Delta I$ and the other multivariate information measures. Namely, the other information measures are concerned with discerning the interactions among the variables in the situation where you know the values of all the variables simultaneously, whereas $\Delta I$ is concerned with comparing that situation to the independent model described by Eq. (\ref{EQ15}) (see Section \ref{Examp4} for further discussion of this topic). 

\begin{table}
\caption{Values of conditional probabilities used to calculate $\Delta I$ for the AND-gate. \label{TableII}}
\begin{tabular}{ r r r|c|c} 
y & $x_1$ & $x_2$ & $p_{ind}(y \mid x_1,x_2)$ & $p(y \mid x_1,x_2)$ \\ \hline 
$0$ & $0$ & $0$ & $1$ & $1$ \\ 
$0$ & $1$ & $0$ & $1$ & $1$ \\ 
$0$ & $0$ & $1$ & $1$ & $1$ \\ 
$0$ & $1$ & $1$ & $0.25$ & $0$ \\ 
$1$ & $0$ & $0$ & $0$ & $0$ \\ 
$1$ & $1$ & $0$ & $0$ & $0$ \\ 
$1$ & $0$ & $1$ & $0$ & $0$ \\ 
$1$ & $1$ & $1$ & $0.75$ & $1$ \\ 
\end{tabular}
\end{table}

The values of the dual total correlation for the XOR-gate example in Table \ref{TableI} demonstrate a crucial difference between the dual total correlation and the other multivariate information measures. Namely, the dual total correlation does not differentiate between the $X$ and $Y$ variables. So, dependencies between all variables are treated equally. In the case of the XOR-gate, the entropy of any variable conditioned on the other two is zero. However, the joint entropy between all variables is 2 bits, so the dual total correlation is equal to 2. Clearly, this result is greater than $I(X_1,X_2;Y)$ for this example. So, if we assume the synergy and redundancy are some portion of $I(X_1,X_2;Y)$, the dual total correlation cannot be the synergy or the redundancy. However, this result is not surprising given the fact that, if we assume the synergy and redundancy are some portion of $I(X_1,X_2;Y)$, the synergy and redundancy require some differentiation between the $X$ variables and $Y$ variables. Since the dual total correlation does not incorporate this distinction, we should expect that it measures a fundamentally different quantity (see Section \ref{Examp6} for further discussion of this topic). 

\subsection{Example 4} \label{Examp4}

Another relevant example for $\Delta I$ is shown in Table \ref{TableIII}. The crucial point to draw from this example is that $\Delta I$ can be greater than $I(X_1,X_2;Y)$. This appears to be in conflict with the intuitive notion of synergy as some part of the information the $X$ variables provide about the $Y$ variable. Why, in this case, $\Delta I$ is greater than $I(X_1,X_2;Y)$ is not immediately clear. To better understand this result, we can examine the difference between $\Delta I$ and $I(X_1,X_2;Y)$. Using Eq. (\ref{EQ18}), (\ref{EQ16}), and (\ref{EQ5}), this difference can be expressed as:
\begin{equation} \label{EQ32}
I(S;Y)-\Delta I(S;Y)=\sum_{\vec{x} \in S, y \in Y}p(y,\vec{x})\log\left(\frac{p_{ind}(\vec{x}|y)}{p_{ind}(\vec{x})}\right)
\end{equation}
The quantity expressed on the RHS of Eq. (\ref{EQ32}), though similar in form, is not a mutual information. Based on the example in Table \ref{TableIII} and the examples in Table \ref{TableI}, this quantity can be positive or negative. Schneidman et. al. further explore this and other noteworthy features of $\Delta I$ \cite{SchneidmanBialek2003}. Fundamentally, $\Delta I$ is a comparison between the complete data and an independent model (as expressed in Eq. (\ref{EQ15})). As Schneidman et. al. note, alternative models could be chosen for the purpose of measuring the importance of correlations between the $X$ variables in the data. We wish to emphasize that $\Delta I$ can provide useful information about a system, but that it measures a fundamentally different quantity in comparison to the other multivariate information measures.

\begin{table}
\caption{Example 4. For this system, $\Delta I$ is greater than $I(X_1,X_2;Y)$. Schneidman et. al. also present an example that demonstrates that $\Delta I$ is not bound by $I(X_1,X_2;Y)$ \cite{SchneidmanBialek2003}. \label{TableIII}}
\begin{tabular}{>{\hfill}ccc|c} 
\multicolumn{3}{c|}{ } & Ex. 4 \\ \hline
$p(x_1,x_2,y)$ & $x_1$ & $x_2$ & $y$ \\ \hline
$1/10$ & $0$ & $0$ & $0$ \\ 
$1/10$ & $1$ & $1$ & $0$ \\ 
$2/10$ & $0$ & $0$ & $1$ \\ 
$6/10$ & $1$ & $1$ & $1$ \\ \hline 
\multicolumn{3}{c|}{$I(X_1;Y)$} & $0.0323$ \\ 
\multicolumn{3}{c|}{$I(X_2;Y)$} & $0.0323$ \\ 
\multicolumn{3}{c|}{$I(X_1,X_2;Y)$} & $0.0323$ \\ 
\multicolumn{3}{c|}{$II(X_1;X_2;Y)$} & $-0.0323$ \\ 
\multicolumn{3}{c|}{$TC(X_1;X_2;Y)$} & $0.9136$ \\ 
\multicolumn{3}{c|}{$DTC(X_1;X_2;Y)$} & $0.8813$ \\ 
\multicolumn{3}{c|}{$\Delta I(X_1,X_2;Y)$} & $0.0337$ \\ 
\multicolumn{3}{c|}{$RSI(X_1,X_2;Y)$} & $-0.0323$ \\ 
\multicolumn{3}{c|}{$VS(X_1,X_2;Y)$} & $-0.0323$ \\ 
\multicolumn{3}{c|}{$\Pi_R(Y;\{1\}\{2\})$} & $0.0323$ \\ 
\multicolumn{3}{c|}{$\Pi_R(Y;\{1\})$} & $0$ \\ 
\multicolumn{3}{c|}{$\Pi_R(Y;\{2\})$} & $0$ \\ 
\multicolumn{3}{c|}{$\Pi_R(Y;\{12\})$} & $0$ \\ 
\end{tabular}
\end{table}

\subsection{Example 5}

The example shown in Table \ref{TableIV} highlights some interesting differences between the information measures, especially regarding the partial information decomposition. Results from the partial information decomposition indicate that 1 bit of information about $Y$ is provided redundantly by $X_1$ and $X_2$, while 1 bit is provided synergistically. This situation is similar to the AND-gate above. Each $X$ variable provides 1 bit of information about $Y$, but both $X$ variables provide the same amount of information about each state of $Y$. So, the partial information decomposition concludes that all of the information is redundant. It should be noted that this is the case despite the fact that \emph{$X_1$ and $X_2$ provide information about different states of $Y$}. $X_1$ can differentiate between $y = 0$ and $y = 2$ on the one hand and $y = 1$ and $y = 3$ on the other, while $X_2$ can differentiate between $y = 0$ and $y = 1$ on the one hand and $y = 2$ and $y = 3$ on the other. Even though the $X$ variables provide information about different states of $Y$, the partial information decomposition is blind to this distinction and concludes, since the $X$ variables provide the same amount of information about each state of $Y$, that their contributions are redundant. Because all of the mutual information between each $X$ variable considered individually is taken up by redundant information, the partial information decomposition concludes there is no unique information and, thus, the remaining 1 bit of information must be synergistic. 

\begin{table}
\caption{Example 5: $Y$ obtains a different state for each unique combination of $X_1$ and $X_2$. The partial information decomposition indicates the presence of redundancy because the $X$ variables provide the same amount of information about each state of $Y$, despite the fact that the $X$ variables provide information about different states of $Y$. The interaction information and $\Delta I$ provide null results. Griffith and Koch also discuss this example in relation to multivariate information measures \cite{Griffith2011}. \label{TableIV}}
\begin{tabular}{>{\hfill}ccc|c} 
\multicolumn{3}{c|}{ } & Ex. 5 \\ \hline
$p(x_1,x_2,y)$ & $x_1$ & $x_2$ & $y$ \\ \hline
$1/4$ & $0$ & $0$ & $0$ \\ 
$1/4$ & $1$ & $0$ & $1$ \\ 
$1/4$ & $0$ & $1$ & $2$ \\ 
$1/4$ & $1$ & $1$ & $3$ \\ \hline 
\multicolumn{3}{c|}{$I(X_1;Y)$} & $1$ \\ 
\multicolumn{3}{c|}{$I(X_2;Y)$} & $1$ \\ 
\multicolumn{3}{c|}{$I(X_1,X_2;Y)$} & $2$ \\ 
\multicolumn{3}{c|}{$II(X_1;X_2;Y)$} & $0$ \\ 
\multicolumn{3}{c|}{$TC(X_1;X_2;Y)$} & $2$ \\ 
\multicolumn{3}{c|}{$DTC(X_1;X_2;Y)$} & $2$ \\ 
\multicolumn{3}{c|}{$\Delta I(X_1,X_2;Y)$} & $0$ \\ 
\multicolumn{3}{c|}{$RSI(X_1,X_2;Y)$} & $0$ \\ 
\multicolumn{3}{c|}{$VS(X_1,X_2;Y)$} & $0$ \\ 
\multicolumn{3}{c|}{$\Pi_R(Y;\{1\}\{2\})$} & $1$ \\ 
\multicolumn{3}{c|}{$\Pi_R(Y;\{1\})$} & $0$ \\ 
\multicolumn{3}{c|}{$\Pi_R(Y;\{2\})$} & $0$ \\ 
\multicolumn{3}{c|}{$\Pi_R(Y;\{12\})$} & $1$ \\ 
\end{tabular}
\end{table}

Example 5 demonstrates the conditions for null results from the interaction information and $\Delta I$. When considering the relationship between one of the $X$ variables and $Y$, we see that knowing the state of the $X$ variable reduces the uncertainty about $Y$ by 1 bit in all cases. However, knowing both $X$ variables only provides 2 bits of information about $Y$. So, the interaction information must be zero because no additional information about $Y$ is gained or lost by knowing both $X$ variables together compared to knowing them each individually. Similarly, $\Delta I$ must be zero because the knowledge of the state of $X_1$ and $X_2$ simultaneously does not provide any additional knowledge about $Y$ compared to the independent models for the relationships between each $X$ variable and $Y$. 

\subsection{Example 6} \label{Examp6}

The example shown in Table \ref{TableV} demonstrates a significant feature of the total correlation. Even when no information is passing to one of the variables considered, the total correlation and the dual total correlation can still produce non-zero results if interactions are present between other variables in the system. This result can be clearly understood using the expression for the total correlation in Eq. (\ref{EQ29}) and the expression for the dual total correlation in Eq. (\ref{EQ33}). The total correlation sums the information passing between variables from the smallest scale (two variables) to the largest scale ($n$ variables). It will detect relationships at all levels and it is unable to differentiate between those levels. The dual total correlation compares the total correlation to the amount of information passing between each individual variable and all other variables considered together as a single vector valued variable. As with the dual total correlation, the total correlation does not differentiate between the $X$ and $Y$ variables, unlike several of the other information measures. 

\begin{table}
\caption{Example 6. All information measures, with the exceptions of the total correlation and the dual total correlation, are zero. The total correlation and the dual total correlation produce non-zero results because they detect interactions between the $X$ variables. \label{TableV}}
\begin{tabular}{>{\hfill}ccc|c} 
\multicolumn{3}{c|}{ } & Ex. 6 \\ \hline
$p(x_1,x_2,y)$ & $x_1$ & $x_2$ & $y$ \\ \hline
$1/2$ & $0$ & $0$ & $0$ \\ 
$1/2$ & $1$ & $1$ & $0$ \\ \hline 
\multicolumn{3}{c|}{$I(X_1;Y)$} & $0$ \\ 
\multicolumn{3}{c|}{$I(X_2;Y)$} & $0$ \\ 
\multicolumn{3}{c|}{$I(X_1,X_2;Y)$} & $0$ \\ 
\multicolumn{3}{c|}{$II(X_1;X_2;Y)$} & $0$ \\ 
\multicolumn{3}{c|}{$TC(X_1;X_2;Y)$} & $1$ \\ 
\multicolumn{3}{c|}{$DTC(X_1;X_2;Y)$} & $1$ \\ 
\multicolumn{3}{c|}{$\Delta I(X_1,X_2;Y)$} & $0$ \\ 
\multicolumn{3}{c|}{$RSI(X_1,X_2;Y)$} & $0$ \\ 
\multicolumn{3}{c|}{$VS(X_1,X_2;Y)$} & $0$ \\ 
\multicolumn{3}{c|}{$\Pi_R(Y;\{1\}\{2\})$} & $0$ \\ 
\multicolumn{3}{c|}{$\Pi_R(Y;\{1\})$} & $0$ \\ 
\multicolumn{3}{c|}{$\Pi_R(Y;\{2\})$} & $0$ \\ 
\multicolumn{3}{c|}{$\Pi_R(Y;\{12\})$} & $0$ \\ 
\end{tabular}
\end{table}

In this case, $Y$ has no entropy, so all information terms that depend on the entropy of $Y$ (i.e., all of the other information measures considered here) are zero. This is expected since all of the other information measures are either explicitly focused on the relationship between the $X$ variables and the $Y$ variable or only focus on interactions that involve all variables. 

\subsection{Examples 7 and 8: three-input Boolean logic gates}

The three-input Boolean logic gate examples shown in Table \ref{TableVI} allow for a comparison between the interaction information, the redundancy-synergy index, Varadan's synergy, and the partial information decomposition. The three-way XOR gate produces similar results to the XOR-gate shown in Table \ref{TableI}. All of the information measures indicate the presence of a synergistic interaction. The partial information decomposition is able to localize the synergy to an interaction between all three $X$ variables. 

Significant differences appear between the information measures when an extraneous $X_3$ variable is added to a basic XOR-gate between $X_1$ and $X_2$. In this case, the interaction information is zero because there is no synergy present between all three $X$ variables. This is despite the fact that the interaction information indicated synergy was present for the basic XOR-gate. Thus, we can see that the interaction information focuses only on interactions between \emph{all} of the $X$ variables and the $Y$ variable. A similar result is observed with Varadan's synergy. Despite the fact that it indicated the presence of synergy in the basic XOR gate, Varadan's synergy does not indicate synergy is present in this logic gate because it also focuses only on interactions between all of the $X$ variables and the $Y$ variable. Both the redundancy-synergy index and the partial information decomposition return results that indicate the presence of synergy between the $X$ variables and the $Y$ variable, but only the partial information decomposition is able to localize the synergy to the $X_1$ and $X_2$ variables. 

\begin{table}
\caption{Examples 7 and 8: three-input Boolean logic gates. All partial information decomposition terms not shown in the table are zero. 3XOR: Three-way XOR-gate. All information measures produce consistent results. $X_1X_2$XOR: XOR-gate involving only $X_1$ and $X_2$. The redundancy-synergy index identifies a synergistic interaction and $\Delta I$ identifies the importance of correlations between the $X$ variables. The partial information decomposition also identifies the variables involved in the synergistic interaction. The interaction information and Varadan's synergy do not identify a synergistic interaction. \label{TableVI}}
\begin{tabular}{>{\hfill}cccc|c|c} 
\multicolumn{4}{c|}{ } & 3XOR & $X_1X_2$XOR \\ \hline
$p(x_1,x_2,y)$ & $x_1$ & $x_2$ & $x_3$ & $y$ & $y$ \\ \hline
$1/8$ & $0$ & $0$ & $0$ & $0$ & $0$ \\ 
$1/8$ & $1$ & $0$ & $0$ & $1$ & $1$ \\ 
$1/8$ & $0$ & $1$ & $0$ & $1$ & $1$ \\ 
$1/8$ & $1$ & $1$ & $0$ & $0$ & $0$ \\ 
$1/8$ & $0$ & $0$ & $1$ & $1$ & $0$ \\ 
$1/8$ & $1$ & $0$ & $1$ & $0$ & $1$ \\ 
$1/8$ & $0$ & $1$ & $1$ & $0$ & $1$ \\ 
$1/8$ & $1$ & $1$ & $1$ & $1$ & $0$ \\ \hline 
\multicolumn{4}{c|}{$I(X_1;Y)$} & $0$ & $0$ \\ 
\multicolumn{4}{c|}{$I(X_2;Y)$} & $0$ & $0$ \\ 
\multicolumn{4}{c|}{$I(X_3;Y)$} & $0$ & $0$ \\ 
\multicolumn{4}{c|}{$I(X_1,X_2;Y)$} & $1$ & $1$ \\ 
\multicolumn{4}{c|}{$II(X_1;X_2;Y)$} & $1$ & $0$ \\ 
\multicolumn{4}{c|}{$TC(X_1;X_2;Y)$} & $1$ & $1$ \\ 
\multicolumn{4}{c|}{$DTC(X_1;X_2;Y)$} & $3$ & $2$ \\ 
\multicolumn{4}{c|}{$\Delta I(X_1,X_2;Y)$} & $1$ & $1$ \\ 
\multicolumn{4}{c|}{$RSI(X_1,X_2;Y)$} & $1$ & $1$ \\ 
\multicolumn{4}{c|}{$VS(X_1,X_2;Y)$} & $1$ & $0$ \\ 
\multicolumn{4}{c|}{$\Pi_R(Y;\{12\})$} & $0$ & $1$ \\ 
\multicolumn{4}{c|}{$\Pi_R(Y;\{123\})$} & $1$ & $0$ \\ 
\end{tabular}
\end{table}

\subsection{Examples 9 to 13: simple model networks}

\begin{figure}
\includegraphics[width=5cm]{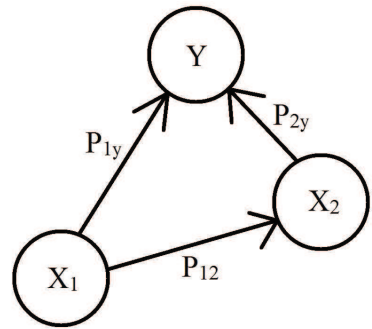}
\caption{Structure of model network used for Examples 9 to 13. \label{Fig1}}
\end{figure}

In an effort to discuss results more directly applicable to several research topics, we will now apply the multivariate information measures to several variations of a simple model network. The general structure of the network is shown in Fig. \ref{Fig1}. The network contains three nodes, each of which can be in one of two states (0 or 1) at any given point in time. The default state of each node is 0. At each time step, there is a certain probability, call it $p_r$, that a given node will be in state 1. The probability that a given node is in state 1 can also be increased if it receives a connection from another node. This driving effect is noted by $p_{1y}$ for the connection from $X_1$ to $Y$, $p_{12}$ for the connection from $X_1$ to $X_2$, and $p_{2y}$ for the connection from $X_2$ to $Y$. All states of the network are determined simultaneously and are independent of the previous states of the network. (See Appendix \ref{Ap3} for further details regarding this model.)

For this simple system, we will discuss five combinations of pr, p1y, p12, and p2y that correspond to noteworthy network topologies. The information theoretic results for these examples are presented in Table \ref{TableVII}. 

\begin{table}
\caption{Examples 9 to 13: simple model network. All information values are in millibits. \label{TableVII}}
\begin{tabular}{>{\hfill}c|c|c|c|c|c} 
 & Ex. 9 & Ex. 10 & Ex. 11 & Ex. 12 & Ex. 13 \\ \hline
\raisebox{-0.5\height} {Diagram} & \raisebox{-0.8\height}{\includegraphics[width=1cm]{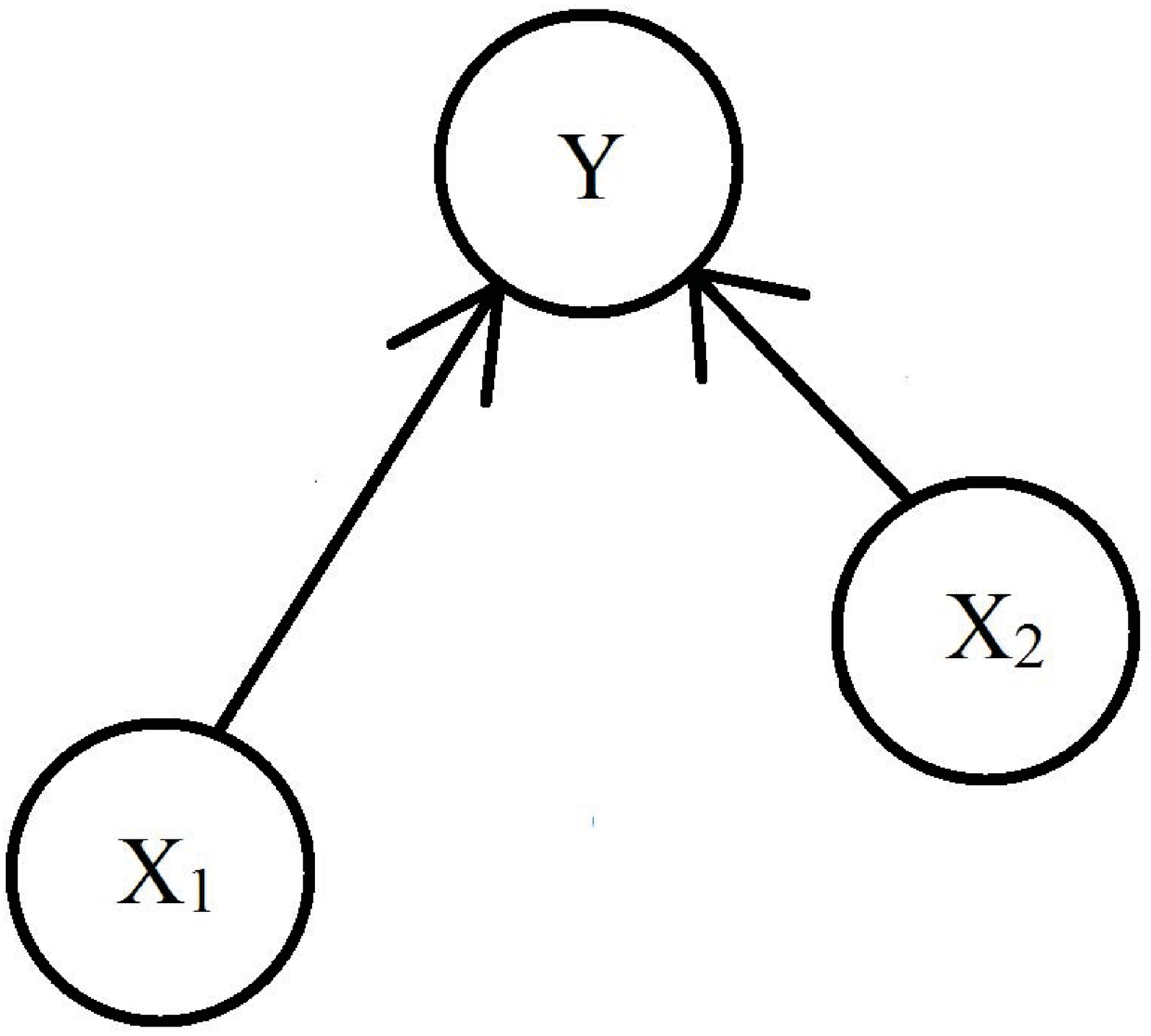}} & \raisebox{-0.8\height}{\includegraphics[width=1cm]{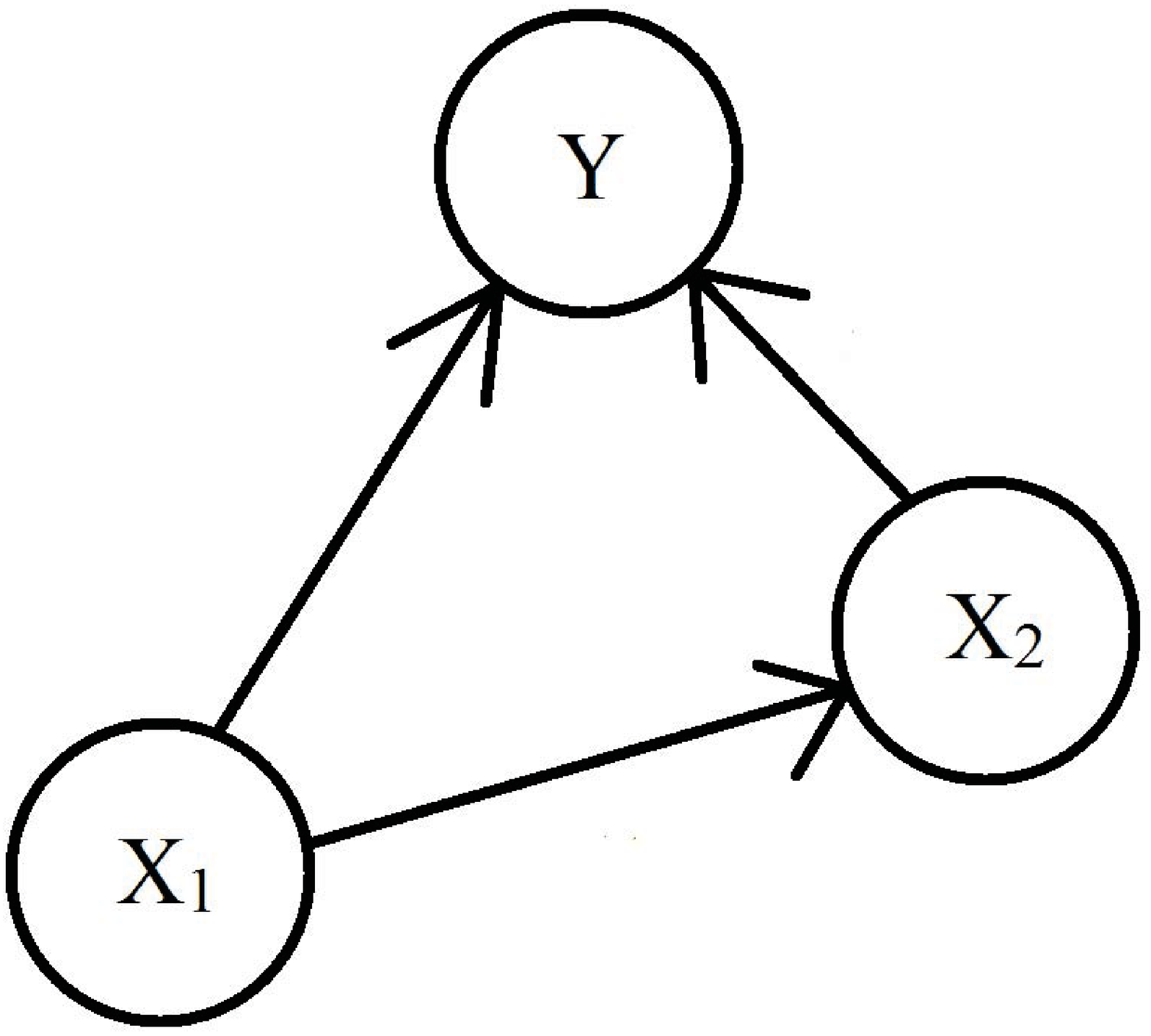}} & \raisebox{-0.8\height}{\includegraphics[width=1cm]{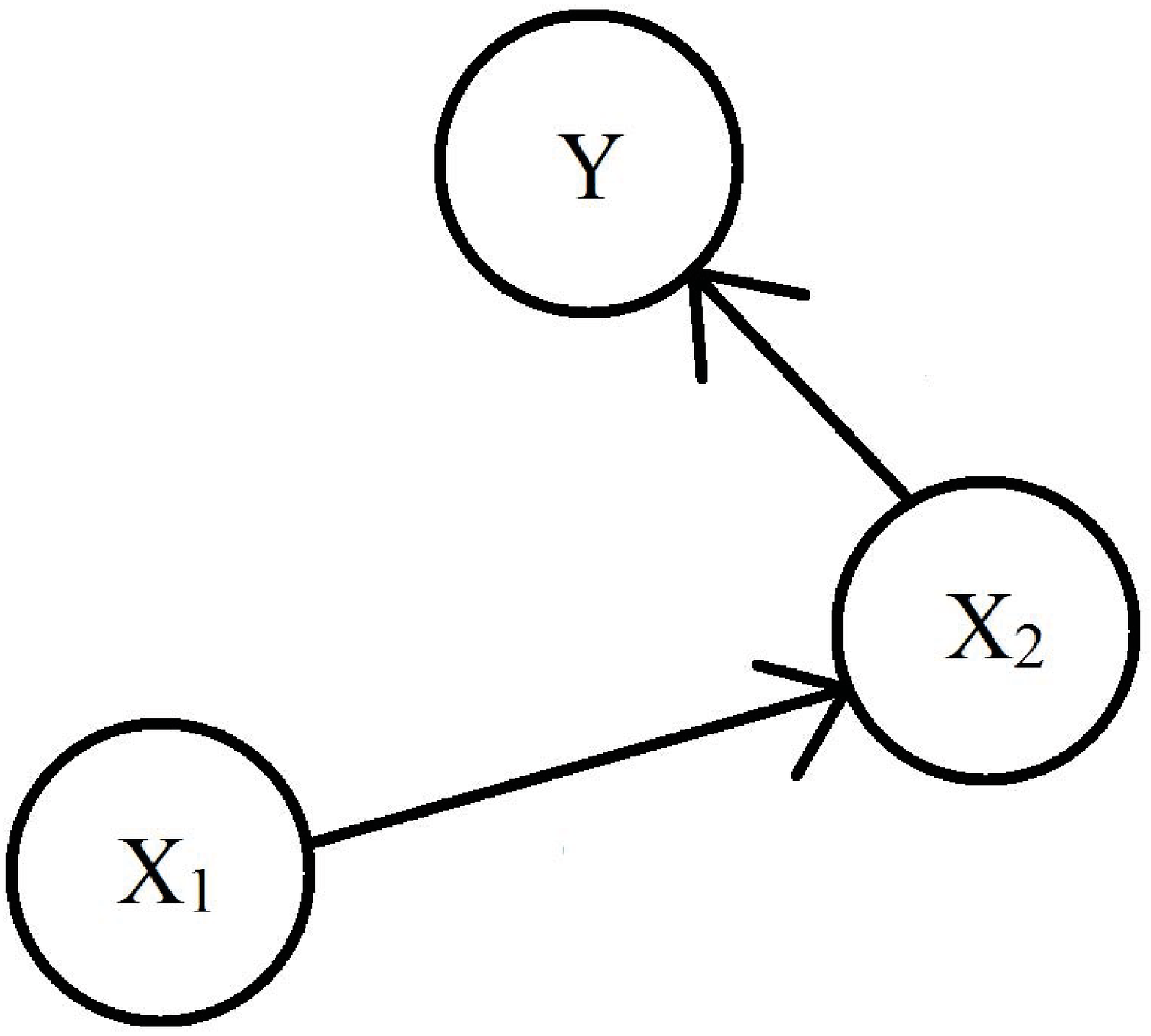}} & \raisebox{-0.8\height}{\includegraphics[width=1cm]{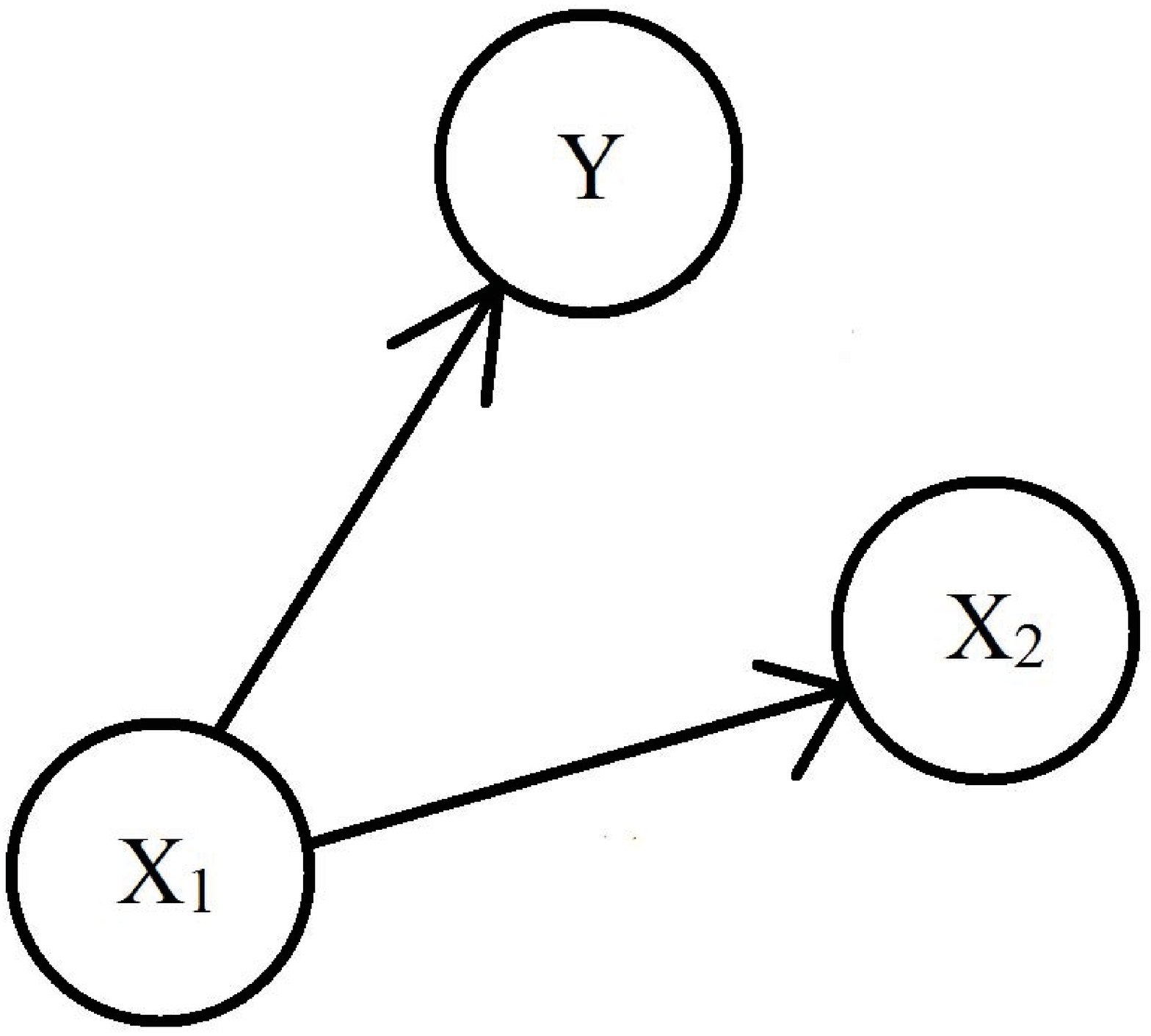}} & \raisebox{-0.8\height}{\includegraphics[width=1cm]{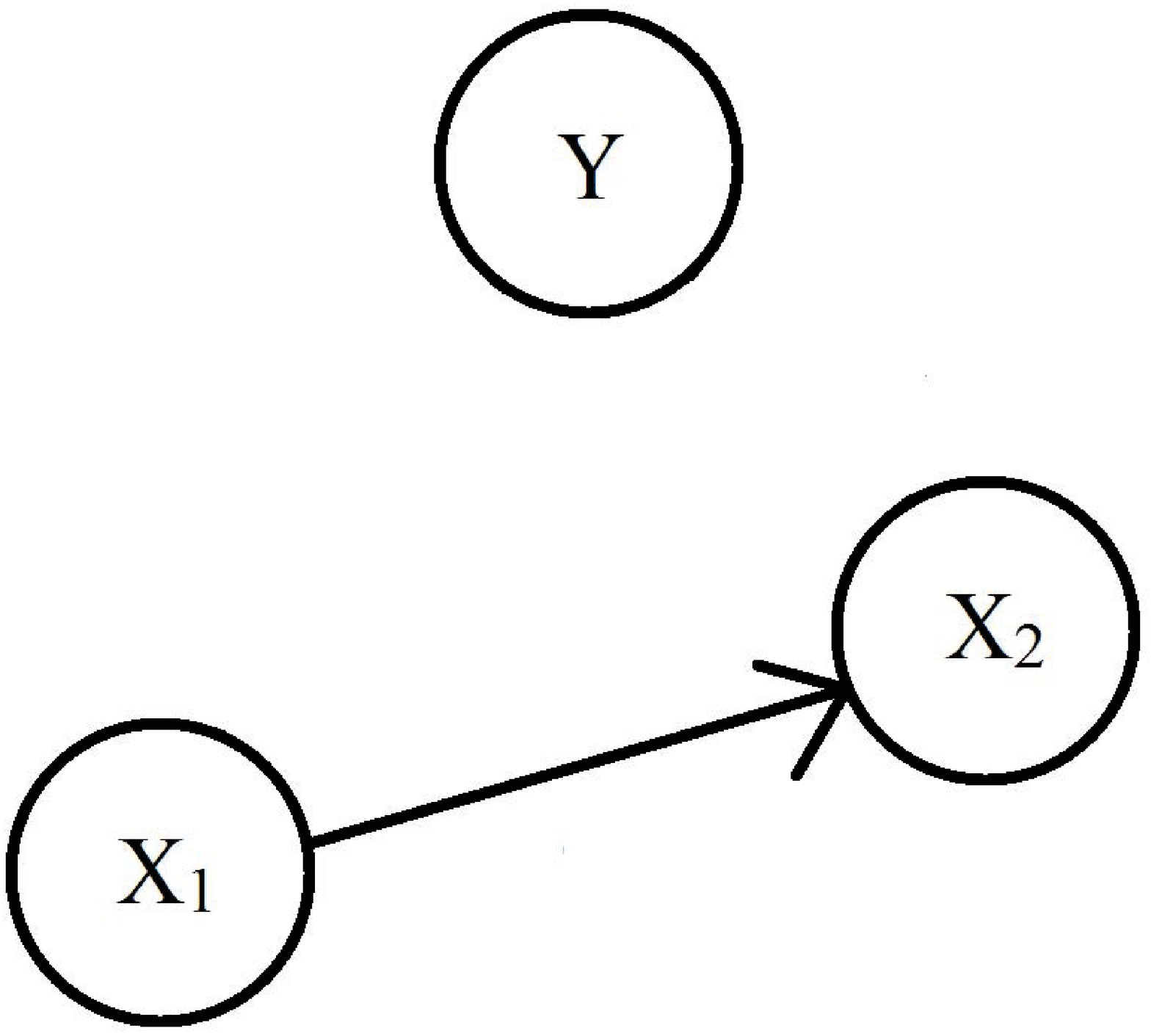}} \\ \hline
$p_r$ & $0.02$ & $0.02$ & $0.02$ & $0.02$ & $0.02$ \\ 
$p_{12}$ & $0$ & $0.1$ & $0.1$ & $0.1$ & $0.1$ \\ 
$p_{1y}$ & $0.1$ & $0.1$ & $0$ & $0.1$ & $0$ \\ 
$p_{2y}$ & $0.1$ & $0.1$ & $0.1$ & $0$ & $0$ \\ \hline 
$I(X_1;Y)$ & $3.061$ & $3.498$ & $0.053$ & $3.225$ & $0$ \\ 
$I(X_2;Y)$ & $3.061$ & $3.801$ & $3.527$ & $0.050$ & $0$ \\ 
$I(X_1,X_2;Y)$ & $6.239$ & $6.750$ & $3.527$ & $3.225$ & $0$ \\ 
$II(X_1;X_2;Y)$ & $0.117$ & $-0.548$ & $-0.053$ & $-0.050$ & $0$ \\ 
$TC(X_1;X_2;Y)$ & $6.239$ & $9.975$ & $6.752$ & $6.450$ & $3.225$ \\ 
$DTC(X_1;X_2;Y)$ & $6.356$ & $9.427$ & $6.698$ & $6.400$ & $3.225$ \\ 
$\Delta I(X_1,X_2;Y)$ & $0.080$ & $0.499$ & $0.064$ & $0.059$ & $0$ \\ 
$RSI(X_1,X_2;Y)$ & $0.117$ & $-0.548$ & $-0.053$ & $-0.050$ & $0$ \\ 
$VS(X_1,X_2;Y)$ & $0.117$ & $-0.548$ & $-0.053$ & $-0.050$ & $0$ \\ 
$\Pi_R(Y;\{1\}\{2\})$ & $3.061$ & $3.498$ & $0.053$ & $0.050$ & $0$ \\ 
$\Pi_R(Y;\{1\})$ & $0$ & $0$ & $0$ & $3.175$ & $0$ \\ 
$\Pi_R(Y;\{2\})$ & $0$ & $0.303$ & $3.473$ & $0$ & $0$ \\ 
$\Pi_R(Y;\{12\})$ & $3.178$ & $2.950$ & $0$ & $0$ & $0$ \\ 
\end{tabular}
\end{table}

Example 9 represents a system where the $X$ nodes independently drive the $Y$ node. Similarly to Example 5, the partial information decomposition indicates that the information from $X_1$ and $X_2$ is entirely redundant and synergistic. This result is somewhat counter intuitive because the $X$ nodes act independently. Again, this is due to the structure of the minimum information in Eq. (\ref{EQ26}). Each $X$ variable provides the same information about each state of $Y$, so the partial information decomposition returns the result that all of the information provided by each $X$ variable about $Y$ is redundant. The interaction information returns a result that indicates the presence of synergy, though the magnitude of this interaction is less than the magnitudes of the synergy and redundancy results from the partial information decomposition. Note that this is the only network for which the interaction information indicates the presence of synergy.

Example 10 is similar to Example 9 with the exception that $X_1$ now also drives $X_2$. Several interesting results are produced for this example. For instance, the total correlation and dual total correlation are significantly elevated in comparison to the other examples. In this example, there is the maximum amount of interactions between all nodes. So, this result agrees with expectations because the total correlation and dual total correlation reflect the total amount of interactions at all scales between all variables. Also, $\Delta I$ obtains its highest value for this example because the actual data and the independent model from Eq. (\ref{EQ15}) are more dissimilar due to the interactions between $X_1$ and $X_2$. Interestingly, the partial information decomposition does not indicate the presence of unique information from $X_1$, despite the fact that $X_1$ is directly influencing $Y$. As with example 9 above, the partial information decomposition returns the result that all of the information $X_1$ provides about $Y$ is redundant. In this case, this result is more intuitive because $X_1$ also drives $X_2$. The interaction information returns a significantly larger magnitude result for this example. This is intuitive given the fact that $X_1$ is driving $X_2$ and that both $X$ variables are driving the $Y$ variable. However, it should be noted that the magnitude of the interaction information is significantly less than the magnitude of the synergy and redundancy from the partial information decomposition. Also, the interaction information result implies the presence of redundancy, unlike Example 10. 

Example 11 represents a common problem case when attempting to infer connectivity based solely on node activity. Node $X_1$ drives $X_2$, which in turn drives $Y$. If the activity of $X_2$ is not known, it would appear that $X_1$ is driving $Y$ directly. The partial information decomposition returns the result that any information provided by $X_1$ about $Y$ is redundant and that the vast majority of the information provided by $X_1$ and $X_2$ about $Y$ is unique information from $X_2$. Both of these results appear to accurately reflect the structure of the network. 

Example 12 also represents a common problem case when determining connectivity. Node $X_1$ drives $Y$ and $X_2$. If the activity of $X_1$ is not known, it would appear that $X_2$ is driving $Y$, when, in fact, no connection exists from $X_2$ to $Y$. Similarly to Example 11, the partial information decomposition identifies the majority of the information from $X_1$ and $X_2$ about $Y$ as unique information from $X_1$ and the remaining information as redundant. Again, these results appear to accurately reflect the structure of the network.

The final example is similar to Example 6 above. In this case, no connections exist from $X_1$ or $X_2$ to $Y$, but $X_1$ drives $X_2$. Almost all of information measures indicate a lack of information transmission. However, the total correlation and the dual total correlation pick up the interaction between $X_1$ and $X_2$. The values of the total correlation and the dual total correlation vary approximately linearly with the number of connections in each network example. This, again, demonstrates the fact that the total correlation and the dual total correlation measure interactions between all variables at all scales. 

\subsection{Analysis of dissociated neural culture}

We will now present the results of applying the information measures discussed above to spiking data from a dissociated neural culture as an illustration of the type of analysis that is possible using these information measures.

The data we chose to analyze are described in Wagenaar et. al. and are freely available online \cite{WagenaarPine2006}. The data contain multiunit spiking activity for each of 60 electrodes in the multielectrode array on which the dissociated neural culture was grown. Specifically, we used data from neural culture 2-2. All details regarding the production and maintenance of the culture can be found in \cite{WagenaarPine2006}. We analyzed recordings from eight points in the development in the culture: days in vitro (DIV) 4, 7, 12, 16, 20, 25, 31, and 33. The DIV 16 recording was 60 minutes long, while all others were 45 minutes long.

For this analysis, the data were binned at 16 ms. The probability distributions necessary for the computation of the information measures were created by examining the spike trains for groups of three non-identical electrodes. For a given group of electrodes, one electrode was labeled the $Y$ electrode, while the other two were labeled the $X_1$ and $X_2$ electrodes. Then, for all time steps in the spike trains, the states of the electrodes (spiking or not spiking) were recorded at time $t$ for the $X_1$ and $X_2$ electrodes and at time $t+1$ for the $Y$ electrode. Next, by counting how many times each state appeared throughout the spike train, the joint probabilities $p(y_{t+1},x_{1,t},x_{2,t})$ were calculated, which were then used to calculate the information measures discussed above. This process was repeated for each group of non-identical electrodes. However, to avoid double counting, groups with swapped $X$ variable assignments were only analyzed once. For instance, the group $X_1$ = electrode 3, $X_2$ = electrode 4, and $Y$ = electrode 5 was analyzed, but the group $X_1$ = electrode 4, $X_2$ = electrode 3, and $Y$ = electrode 5 was not analyzed. In order to compensate for the changing firing rate through development of the cultures, all information values for a given group were normalized by the entropy of the Y electrode. 

To illustrate the statistical significance of the information measure values, we also created and analyzed a randomized data set from the original neural culture data. The randomization was accomplished by splitting each electrode spike train at a randomly chosen point and swapping the two remaining pieces. By doing this, the structure of the electrode spike train is almost entirely preserved, but the temporal relationship between the electrode spike trains is significantly disrupted. 

\begin{figure*}
\includegraphics[width=\textwidth]{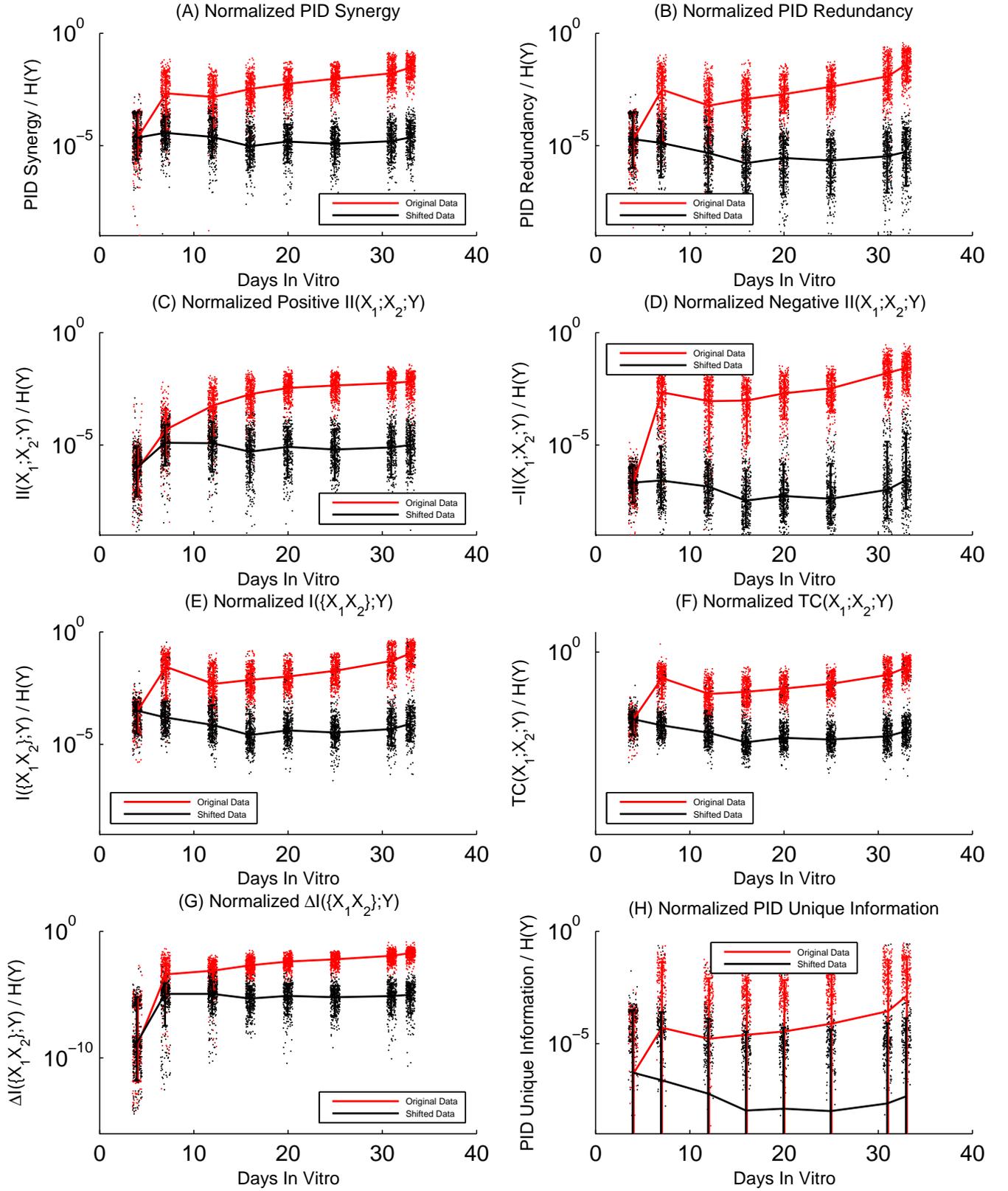}
\caption{Many information measures show changes over neural development. A) PID Synergy. B) PID Redundancy. C) Positive Interaction Information values. D) Negative Interaction Information values. E) Mutual Information. F) Total Correlation. G) $\Delta I$. H) PID Unique Information. All information values are normalized by the entropy of the Y electrode. Each individual data point represents one group of electrodes. To improve clarity, the data points are jittered randomly around the DIV and only 0.4\% of the data points are shown. The line plots show the 90th percentile, median, and 10th percentile of all the data for a given DIV. Note that as the culture matured, the total amount of information transmitted increased and the types of interactions present in the network changed. \label{Fig2}}
\end{figure*}

\begin{figure}
\includegraphics[width=8cm]{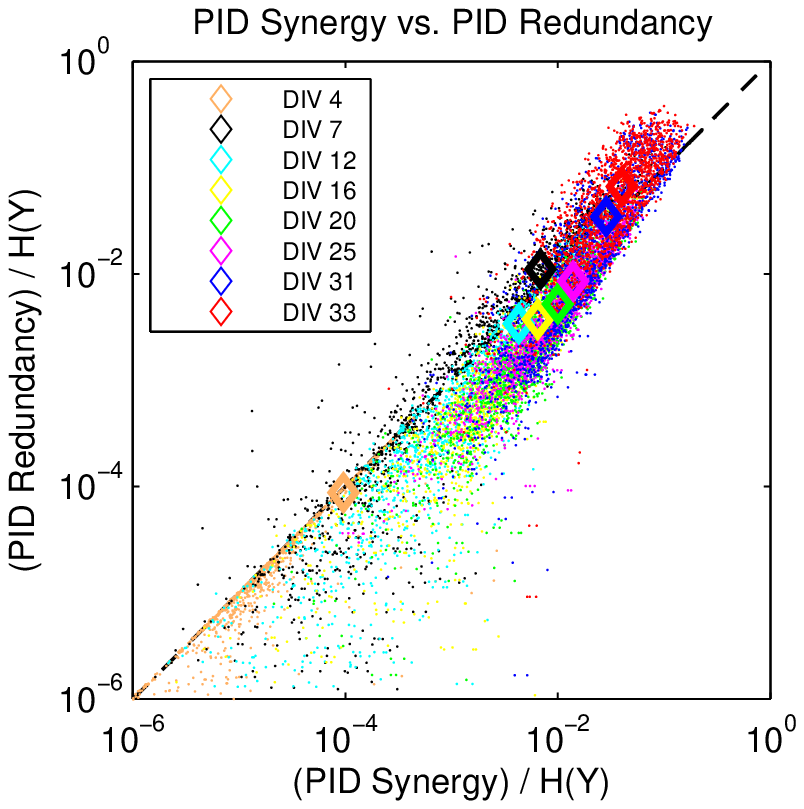}
\caption{The balance of PID synergy and PID redundancy changed during development. Distribution of normalized PID synergy and PID redundancy. Each data point represents the information values for one group of electrodes (only 2\% of the data are shown to improve clarity). Diamonds represent mean values for a given DIV.  \label{Fig3}}
\end{figure}

The results of these analyses are presented in Fig. \ref{Fig2} and Fig. \ref{Fig3}. The results shown in Fig. \ref{Fig2} indicate that at day 4 essentially no information was being transmitted in the network. However, by day 7, a great deal of information was being transmitted, as can be seen by the peaks in mutual information (Fig. \ref{Fig2} E) and the total correlation (Fig. \ref{Fig2} F). As the culture continued to develop after day 7, most information measures decreased and then slowly increased to maxima on the last day, DIV 33. Interestingly, $\Delta I$ (Fig. \ref{Fig2} G) showed an increase at day 7, but then a steady increase afterwards. The total correlation (Fig. \ref{Fig2} F) mimics the changes in the mutual information (Fig. \ref{Fig2} E), but because the total correlation measures the total amount of information being transmitted among the $X$ and $Y$ variables, it possessed higher values than the mutual information.

The relationship between the interaction information and the partial information decomposition was also illustrated through development. As the culture developed, the PID synergy was larger than the PID redundancy. Then, between days 31 and 33, the PID synergy became significantly smaller than the PID redundancy. In the interaction information, this relationship was expressed by positive values through most of the culture's development, with the exception of large negative values at days 7 and 33. However, notice that groups of electrodes with positive and negative interaction information values were found in each recording. To further investigate this relationship, we plotted the distribution of PID synergy and PID redundancy for groups of electrodes (Fig. \ref{Fig3}). This plot shows that the network contained groups of electrodes with slightly more PID redundancy than PID synergy at day 7, but that, after that point, the total amount of information decreased and became more biased towards PID synergy at day 12. From that point, the total amount of information increased up to the last recording where the network was once again biased towards PID redundancy. So, we can relate the results from the partial information decomposition and the interaction information using Fig. \ref{Fig3} by noting that, while the PID synergy and PID redundancy for a given group of electrodes determines a point’s position in Fig. \ref{Fig3}, the interaction information describes how far that point is from the equilibrium line. Given the fact that many points in Fig. \ref{Fig3} are near the equilibrium line, the partial information decomposition finds that many groups of electrodes contain synergistic and redundant interactions simultaneously. This feature would be lost by only examining the interaction information.

Obviously, this analysis could be made significantly more complex and interesting. For instance, the analysis could be improved by including more data sets, varying the variable assignments, using different bin sizes, using more robust methods to test statistical significance, and so forth. However, based on this simple illustration, we believe that it is clear that the information analysis methods discussed herein could be used to address interesting questions related to this system, or other systems. For instance, it may be possible to relate these changes through development to previous work on changes in dissociated cultures through development \cite{Kamioka1996,WagenaarPine2006,WagenaarNadasdy2006,Pasquale2008,Tetzlaff2010}. 

\section{Discussion}

Based on the results from several simple systems, we were able to explore the properties of the multivariate information measures discussed in this paper. We will now discuss each measure in turn.

The oldest multivariate information measure - the interaction information - was shown to focus on interactions between all $X$ variables and the $Y$ variable using the three-input Boolean logic gate examples. Furthermore, the two-input AND-gate demonstrated how the interaction information is related to the excess information provided by both $X$ variables about the $Y$ variable beyond the total amount of information those $X$ variables provide about $Y$ when considered individually. Also, that example demonstrated the relationship between the interaction information and the partial information decomposition as shown in Eq. (\ref{EQ27}). For the model network examples, the interaction information had its largest magnitude when the interactions were present between all three nodes. Also, for these examples, the interaction information indicated the presence of synergy when both nodes $X_1$ and $X_2$ drove $Y$, but not each other (Example 9), while it indicated the presence of redundancy when either node $X_1$ or $X_2$ drove $Y$ and $X_1$ drove $Y$ (Examples 10 to 12). When the interaction information was applied to data from a developing neural culture, it showed changes in the type of interactions present in the network during development. 

In contrast to the interaction information, the total correlation was shown to sum interactions among all variables at all scales using Example 6. In other words, the value of the total correlation for any system incorporates interactions between groups of variables at all scales. This feature was made apparent using the model network examples. There, the total correlation varied approximately linearly with the number of connections present in the network. Furthermore, the total correlation is symmetric with regard to all variables considered, whereas the other information measures focus on the relationship between the set of $X$ variables and the $Y$ variable. When applied to the data from the neural culture, the total correlation and the mutual information both showed increases in the total amount of information being transmitted in the network through development. 

The dual total correlation was found to be similar to the total correlation in that both do not differentiate between the $X$ and $Y$ variables. Also, like the total correlation, the dual total correlation varied approximately linearly with the number of connections in the model network examples. The function of the dual total correlation was also highlighted with the XOR-gate example. There, we saw that the dual total correlation compares the uncertainty with regards to all variables with the total uncertainty that remains about each variable if all the other variables are known.  

Using the AND-gate, $\Delta I$ was shown to measure a subtly different quantity compared to the other information measures. The other information measures seek to evaluate the interactions between the $X$ variables and the $Y$ variable given that one knows the values of all variables simultaneously (i.e. in the case that the total joint probability distribution is known). $\Delta I$ compares that situation to a model where it is assumed that the $X$ variables act independently of one another in an effort to measure the importance of knowing the correlations between the $X$ variables. Clearly, this goal is similar to the goals of the other information measures. However, given the fact that $\Delta I$ can be greater than $I(X_1,X_2;Y)$, as was shown in Example 4, and if we assume the synergy and redundancy are some portion of $I(X_1,X_2;Y)$, $\Delta I$ cannot be the synergy or the redundancy. $\Delta I$ can provide useful information about a system, but the distinction between the structure of $\Delta I$ and the other information measures, along with the fact that $\Delta I$ cannot be the synergy or redundancy as previously defined, should be considered when choosing the appropriate information measure with which to perform an analysis. Unlike several of the other information measures which showed changes in the types of interactions present in the developing neural culture, $\Delta I$ showed a uniform increase in the importance of correlations in the network throughout development. 

The redundancy-synergy index and Varadan's synergy are identical to the interaction information when only two $X$ variables are considered. However, when we examined three-input Boolean logic gates, we found that Varadan's synergy - like the interaction information - was unable to detect a synergistic interaction among a subset of the $X$ variables and the $Y$ variable. The redundancy-synergy index was able to detect this synergy, but it was unable to localize the subset of $X$ variables involved in the interaction.

The partial information decomposition provided interesting and possibly useful results for several of the example systems. When applied to the Boolean logic gates, the partial information decomposition was able to identify the $X$ variables involved in the interactions, unlike all other information measures. Using the AND-gate example, we saw that the partial information decomposition found that both synergy and redundancy were present in the system, unlike the interaction information, which indicated only synergy was present. Perhaps the most illuminating example system for the partial information decomposition was Example 5. In that case, the partial information decomposition concluded that each $X$ variable provided entirely redundant information because each $X$ variable provided the same amount of information about each state of $Y$, even though each $X$ variable provided information about different states of $Y$. This point highlights how the partial information decomposition defines redundancy via Eq. (\ref{EQ26}). It calculates the redundant contributions based only on the quantity of information each $X$ variable provides about each state of $Y$. In the developing neural culture, the partial information decomposition, similar to the interaction information, showed a changing balance between synergy and redundancy through development. However, unlike the interaction information, the partial information decomposition was able to separate simultaneous synergistic and redundant interactions. 

\section{Conclusion}

We applied several multivariate information measures to simple example systems in an attempt to explore the properties of the information measures. We found that the information measures produce similar or identical results for some systems (e.g. XOR-gate), but that the measures produce different results for other systems. In examining these results, we found several subtle differences between the information measures that impacted the results. Based on the understanding gained from these simple systems, we were able to apply the information measures to spiking data from a neural culture through its development. Based on this illustrative analysis, we saw interesting changes in the amount of information being transmitted and the interactions present in the network. 

We wish to emphasize that none of these information measures is the ``right'' measure. All of them produce results that can be used to learn something about the system being studied. We hope that this work will assist other researchers as they deliberate on the specific questions they wish to answer about a given system so that they may use the multivariate information measures that best suit their goals. 

\begin{acknowledgements}
We would like to thank Paul Williams, Randy Beer, Alexander Murphy-Nakhnikian, Shinya Ito, Ben Nicholson, Emily Miller, Virgil Griffith, and Elizabeth Timme for their helpful comments on this paper. 
\end{acknowledgements}

\appendix

\section{Additional total correlation derivation}
\label{Ap1}

Eq. (\ref{EQ14}) can be rewritten as Eq. (\ref{EQ29}) by adding and subtracting several joint entropy terms and then using Eq. (\ref{EQ2}). For instance, when $n=3$, we have:
\begin{eqnarray}\label{EQA1}
TC(S) =\left(\sum_{X_i \in S}H(X_i)\right)-H(S) = \hspace{1cm} \nonumber \\
H(X_1)+H(X_2)+H(X_3)-H(X_1,X_2,X_3) = \hspace{0.2cm} \nonumber \\
H(X_1)+H(X_2)-H(X_1,X_2)+H(X_1,X_2)+ \hspace{0.2cm} \nonumber \\
H(X_3)-H(X_1,X_2,X_3) = \hspace{2cm} \nonumber \\
I(X_1;X_2)+I(X_1,X_2;X_3) \hspace{3cm}
\end{eqnarray}
\vspace{1pc}

A similar substitution can be peformed for $n>3$. 

\section{Additional dual total correlation derivation}
\label{Ap2}

Eq. (\ref{EQ30}) can be rewritten as Eq. (\ref{EQ33}) by substituting the expression for the total correlation in Eq. (\ref{EQ14}) and then applying Eq. (\ref{EQ2}). 
\begin{eqnarray}\label{EQB1}
DTC(S)=\left(\sum_{X_i \in S}H(S/X_i)\right)-(n-1)H(S)) = \hspace{0.5cm} \nonumber \\
\left(\sum_{X_i \in S}H(S/X_i)+H(X_i)\right)-nH(S)-TC(S)= \hspace{0.3cm} \nonumber \\
\left(\sum_{X_i \in S}I(S/X_i;X_i)\right)-TC(S) \hspace{1cm}
\end{eqnarray}

\section{Model Network}
\label{Ap3}

Given values for $p_r$, $p_{1y}$, $p_{12}$, and $p_{2y}$, the joint probabilities for all possible states of the network can be calculated. For example:

\begin{equation} \label{EQ34}
p(x_1=1)=p_r
\end{equation}

\begin{equation} \label{EQ35}
p(x_1=0)=1-p_r
\end{equation}

\begin{equation} \label{EQ36}
p(x_2=1|x_1=1)=p_r+p_{12}-p_rp_{12}
\end{equation}

\begin{equation} \label{EQ37}
p(x_2=0|x_1=1)=1-p(x_2=1|x_1=1)
\end{equation}

The joint probabilities for the examples discussed in the main text of the article are shown in Table \ref{TableAI}.

\begin{table}
\caption{Joint probabilities for examples 9 to 13. \label{TableAI}}
\begin{tabular}{>{\hfill}cccccccc} 
\multicolumn{3}{c}{} & Ex. 9 & Ex. 10 & Ex. 11 & Ex. 12 & Ex. 13 \\ \hline
\multicolumn{3}{c}{Diagram} & \raisebox{-0.8\height}{\includegraphics[width=1cm]{Figure2}} & \raisebox{-0.8\height}{\includegraphics[width=1cm]{Figure3}} & \raisebox{-0.8\height}{\includegraphics[width=1cm]{Figure4}} & \raisebox{-0.8\height}{\includegraphics[width=1cm]{Figure5}} & \raisebox{-0.8\height}{\includegraphics[width=1cm]{Figure6}} \\ \hline
\multicolumn{3}{c}{$p_r$} & $0.02$ & $0.02$ & $0.02$ & $0.02$ & $0.02$ \\ 
\multicolumn{3}{c}{$p_{12}$} & $0$ & $0.1$ & $0.1$ & $0.1$ & $0.1$ \\ 
\multicolumn{3}{c}{$p_{1y}$} & $0.1$ & $0.1$ & $0$ & $0.1$ & $0$ \\ 
\multicolumn{3}{c}{$p_{2y}$} & $0.1$ & $0.1$ & $0.1$ & $0$ & $0$ \\ \hline
$x_1$ & $x_2$ & $y$ & \multicolumn{5}{c}{$p(x_1,x_2,y)$} \\ \hline
0 & 0 & 0 & $0.9412$ & $0.9412$ & $0.9412$ & $0.9412$ & $0.9412$ \\ 
1 & 0 & 0 & $0.0173$ & $0.0156$ & $0.0173$ & $0.0156$ & $0.0173$ \\ 
0 & 1 & 0 & $0.0173$ & $0.0173$ & $0.0173$ & $0.0192$ & $0.0192$ \\ 
1 & 1 & 0 & $0.0003$ & $0.0019$ & $0.0021$ & $0.0021$ & $0.0023$ \\ 
0 & 0 & 1 & $0.0192$ & $0.0192$ & $0.0192$ & $0.0192$ & $0.0192$ \\ 
1 & 0 & 1 & $0.0023$ & $0.0021$ & $0.0004$ & $0.0021$ & $0.0004$ \\ 
0 & 1 & 1 & $0.0023$ & $0.0023$ & $0.0023$ & $0.0004$ & $0.0004$ \\ 
1 & 1 & 1 & $0.0001$ & $0.0005$ & $0.0003$ & $0.0003$ & $0.0000$ \\ 
\end{tabular}
\end{table}

\bibliography{RefsManuscript}

\providecommand{\noopsort}[1]{}\providecommand{\singleletter}[1]{#1}%
\begin{thebibliography}{46}%
\makeatletter
\providecommand \@ifxundefined [1]{%
 \@ifx{#1\undefined}
}%
\providecommand \@ifnum [1]{%
 \ifnum #1\expandafter \@firstoftwo
 \else \expandafter \@secondoftwo
 \fi
}%
\providecommand \@ifx [1]{%
 \ifx #1\expandafter \@firstoftwo
 \else \expandafter \@secondoftwo
 \fi
}%
\providecommand \natexlab [1]{#1}%
\providecommand \enquote  [1]{``#1''}%
\providecommand \bibnamefont  [1]{#1}%
\providecommand \bibfnamefont [1]{#1}%
\providecommand \citenamefont [1]{#1}%
\providecommand \href@noop [0]{\@secondoftwo}%
\providecommand \href [0]{\begingroup \@sanitize@url \@href}%
\providecommand \@href[1]{\@@startlink{#1}\@@href}%
\providecommand \@@href[1]{\endgroup#1\@@endlink}%
\providecommand \@sanitize@url [0]{\catcode `\\12\catcode `\$12\catcode
  `\&12\catcode `\#12\catcode `\^12\catcode `\_12\catcode `\%12\relax}%
\providecommand \@@startlink[1]{}%
\providecommand \@@endlink[0]{}%
\providecommand \url  [0]{\begingroup\@sanitize@url \@url }%
\providecommand \@url [1]{\endgroup\@href {#1}{\urlprefix }}%
\providecommand \urlprefix  [0]{URL }%
\providecommand \Eprint [0]{\href }%
\providecommand \doibase [0]{http://dx.doi.org/}%
\providecommand \selectlanguage [0]{\@gobble}%
\providecommand \bibinfo  [0]{\@secondoftwo}%
\providecommand \bibfield  [0]{\@secondoftwo}%
\providecommand \translation [1]{[#1]}%
\providecommand \BibitemOpen [0]{}%
\providecommand \bibitemStop [0]{}%
\providecommand \bibitemNoStop [0]{.\EOS\space}%
\providecommand \EOS [0]{\spacefactor3000\relax}%
\providecommand \BibitemShut  [1]{\csname bibitem#1\endcsname}%
\let\auto@bib@innerbib\@empty
\bibitem [{\citenamefont {Rieke}\ \emph {et~al.}(1997)\citenamefont {Rieke},
  \citenamefont {Warland}, \citenamefont {de~Ruyter~van Steveninck},\ and\
  \citenamefont {Bialek}}]{Rieke1997}%
  \BibitemOpen
  \bibfield  {author} {\bibinfo {author} {\bibfnamefont {F.}~\bibnamefont
  {Rieke}}, \bibinfo {author} {\bibfnamefont {D.}~\bibnamefont {Warland}},
  \bibinfo {author} {\bibfnamefont {R.~R.}\ \bibnamefont {de~Ruyter~van
  Steveninck}}, \ and\ \bibinfo {author} {\bibfnamefont {W.}~\bibnamefont
  {Bialek}},\ }\href@noop {} {\emph {\bibinfo {title} {Spikes: Exploring the
  Neural Code}}}\ (\bibinfo  {publisher} {MIT Press},\ \bibinfo {year}
  {1997})\BibitemShut {NoStop}%
\bibitem [{\citenamefont {Ziv}\ and\ \citenamefont {Lempel}(1977)}]{Ziv1977}%
  \BibitemOpen
  \bibfield  {author} {\bibinfo {author} {\bibfnamefont {J.}~\bibnamefont
  {Ziv}}\ and\ \bibinfo {author} {\bibfnamefont {A.}~\bibnamefont {Lempel}},\
  }\href@noop {} {\bibfield  {journal} {\bibinfo  {journal} {IEEE Trans. Inf.
  Theory}\ }\textbf {\bibinfo {volume} {23}},\ \bibinfo {pages} {337} (\bibinfo
  {year} {1977})}\BibitemShut {NoStop}%
\bibitem [{\citenamefont {Berrou}\ \emph {et~al.}(1993)\citenamefont {Berrou},
  \citenamefont {Glavieux},\ and\ \citenamefont {Thitimajshima}}]{Berrou1993}%
  \BibitemOpen
  \bibfield  {author} {\bibinfo {author} {\bibfnamefont {C.}~\bibnamefont
  {Berrou}}, \bibinfo {author} {\bibfnamefont {A.}~\bibnamefont {Glavieux}}, \
  and\ \bibinfo {author} {\bibfnamefont {P.}~\bibnamefont {Thitimajshima}},\
  }in\ \href@noop {} {\emph {\bibinfo {booktitle} {Proceedings of IEEE
  International Conference on Communications}}},\ Vol.~\bibinfo {volume} {2}\
  (\bibinfo {year} {1993})\ p.\ \bibinfo {pages} {1064}\BibitemShut {NoStop}%
\bibitem [{\citenamefont {Fraser}\ and\ \citenamefont
  {Swinney}(1986)}]{Fraser1986}%
  \BibitemOpen
  \bibfield  {author} {\bibinfo {author} {\bibfnamefont {A.~M.}\ \bibnamefont
  {Fraser}}\ and\ \bibinfo {author} {\bibfnamefont {H.~L.}\ \bibnamefont
  {Swinney}},\ }\href@noop {} {\bibfield  {journal} {\bibinfo  {journal} {Phys.
  Rev. A}\ }\textbf {\bibinfo {volume} {33}},\ \bibinfo {pages} {1134}
  (\bibinfo {year} {1986})}\BibitemShut {NoStop}%
\bibitem [{\citenamefont {Butte}\ and\ \citenamefont
  {Kohane}(2000)}]{Butte2000}%
  \BibitemOpen
  \bibfield  {author} {\bibinfo {author} {\bibfnamefont {A.~J.}\ \bibnamefont
  {Butte}}\ and\ \bibinfo {author} {\bibfnamefont {I.~S.}\ \bibnamefont
  {Kohane}},\ }in\ \href@noop {} {\emph {\bibinfo {booktitle} {Pacific
  Symposium on Biocomputing}}},\ Vol.~\bibinfo {volume} {5}\ (\bibinfo {year}
  {2000})\ p.\ \bibinfo {pages} {415}\BibitemShut {NoStop}%
\bibitem [{\citenamefont {McGill}(1954)}]{McGill1954}%
  \BibitemOpen
  \bibfield  {author} {\bibinfo {author} {\bibfnamefont {W.~J.}\ \bibnamefont
  {McGill}},\ }\href@noop {} {\bibfield  {journal} {\bibinfo  {journal}
  {Psychometrika}\ }\textbf {\bibinfo {volume} {19}},\ \bibinfo {pages} {97}
  (\bibinfo {year} {1954})}\BibitemShut {NoStop}%
\bibitem [{\citenamefont {Watanabe}(1960)}]{Watanabe1960}%
  \BibitemOpen
  \bibfield  {author} {\bibinfo {author} {\bibfnamefont {S.}~\bibnamefont
  {Watanabe}},\ }\href@noop {} {\bibfield  {journal} {\bibinfo  {journal} {IBM
  JJ. Res. Dev.}\ }\textbf {\bibinfo {volume} {4}},\ \bibinfo {pages} {66}
  (\bibinfo {year} {1960})}\BibitemShut {NoStop}%
\bibitem [{\citenamefont {Han}(1975)}]{Han1975}%
  \BibitemOpen
  \bibfield  {author} {\bibinfo {author} {\bibfnamefont {T.~S.}\ \bibnamefont
  {Han}},\ }\href@noop {} {\bibfield  {journal} {\bibinfo  {journal}
  {Information and Control}\ }\textbf {\bibinfo {volume} {29}},\ \bibinfo
  {pages} {337} (\bibinfo {year} {1975})}\BibitemShut {NoStop}%
\bibitem [{\citenamefont {Chechik}\ \emph {et~al.}(2001)\citenamefont
  {Chechik}, \citenamefont {Globerson}, \citenamefont {Tishby}, \citenamefont
  {Anderson}, \citenamefont {Young},\ and\ \citenamefont
  {Nelken}}]{Chechik2001}%
  \BibitemOpen
  \bibfield  {author} {\bibinfo {author} {\bibfnamefont {G.}~\bibnamefont
  {Chechik}}, \bibinfo {author} {\bibfnamefont {A.}~\bibnamefont {Globerson}},
  \bibinfo {author} {\bibfnamefont {N.}~\bibnamefont {Tishby}}, \bibinfo
  {author} {\bibfnamefont {M.~J.}\ \bibnamefont {Anderson}}, \bibinfo {author}
  {\bibfnamefont {E.~D.}\ \bibnamefont {Young}}, \ and\ \bibinfo {author}
  {\bibfnamefont {I.}~\bibnamefont {Nelken}},\ }in\ \href@noop {} {\emph
  {\bibinfo {booktitle} {Neaural Information Processing Systems 14}}},\
  Vol.~\bibinfo {volume} {1},\ \bibinfo {editor} {edited by\ \bibinfo {editor}
  {\bibfnamefont {T.~G.}\ \bibnamefont {Dietterich}}, \bibinfo {editor}
  {\bibfnamefont {S.}~\bibnamefont {Becker}}, \ and\ \bibinfo {editor}
  {\bibfnamefont {Z.}~\bibnamefont {Ghahramani}}}\ (\bibinfo  {publisher} {MIT
  Press},\ \bibinfo {year} {2001})\ p.\ \bibinfo {pages} {173}\BibitemShut
  {NoStop}%
\bibitem [{\citenamefont {Nirenberg}\ \emph {et~al.}(2001)\citenamefont
  {Nirenberg}, \citenamefont {Carcieri}, \citenamefont {Jacobs},\ and\
  \citenamefont {Latham}}]{Nirenberg2001}%
  \BibitemOpen
  \bibfield  {author} {\bibinfo {author} {\bibfnamefont {S.}~\bibnamefont
  {Nirenberg}}, \bibinfo {author} {\bibfnamefont {S.~M.}\ \bibnamefont
  {Carcieri}}, \bibinfo {author} {\bibfnamefont {A.~L.}\ \bibnamefont
  {Jacobs}}, \ and\ \bibinfo {author} {\bibfnamefont {P.~E.}\ \bibnamefont
  {Latham}},\ }\href@noop {} {\bibfield  {journal} {\bibinfo  {journal}
  {Nature}\ }\textbf {\bibinfo {volume} {411}},\ \bibinfo {pages} {698}
  (\bibinfo {year} {2001})}\BibitemShut {NoStop}%
\bibitem [{\citenamefont {Schneidman}\ \emph
  {et~al.}(2003{\natexlab{a}})\citenamefont {Schneidman}, \citenamefont
  {Still}, \citenamefont {{Berry II}},\ and\ \citenamefont
  {Bialek}}]{SchneidmanStill2003}%
  \BibitemOpen
  \bibfield  {author} {\bibinfo {author} {\bibfnamefont {E.}~\bibnamefont
  {Schneidman}}, \bibinfo {author} {\bibfnamefont {S.}~\bibnamefont {Still}},
  \bibinfo {author} {\bibfnamefont {M.~J.}\ \bibnamefont {{Berry II}}}, \ and\
  \bibinfo {author} {\bibfnamefont {W.}~\bibnamefont {Bialek}},\ }\href@noop {}
  {\bibfield  {journal} {\bibinfo  {journal} {Physical Review Letters}\
  }\textbf {\bibinfo {volume} {91}},\ \bibinfo {pages} {238701} (\bibinfo
  {year} {2003}{\natexlab{a}})}\BibitemShut {NoStop}%
\bibitem [{\citenamefont {Varadan}\ \emph {et~al.}(2006)\citenamefont
  {Varadan}, \citenamefont {III},\ and\ \citenamefont
  {Anastassiou}}]{Varadan2006}%
  \BibitemOpen
  \bibfield  {author} {\bibinfo {author} {\bibfnamefont {V.}~\bibnamefont
  {Varadan}}, \bibinfo {author} {\bibfnamefont {D.~M.~M.}\ \bibnamefont {III}},
  \ and\ \bibinfo {author} {\bibfnamefont {D.}~\bibnamefont {Anastassiou}},\
  }\href@noop {} {\bibfield  {journal} {\bibinfo  {journal} {Bioinformatics}\
  }\textbf {\bibinfo {volume} {22}},\ \bibinfo {pages} {e497} (\bibinfo {year}
  {2006})}\BibitemShut {NoStop}%
\bibitem [{\citenamefont {Williams}\ and\ \citenamefont
  {Beer}(2010)}]{Williams2010}%
  \BibitemOpen
  \bibfield  {author} {\bibinfo {author} {\bibfnamefont {P.~L.}\ \bibnamefont
  {Williams}}\ and\ \bibinfo {author} {\bibfnamefont {R.~D.}\ \bibnamefont
  {Beer}},\ }\href@noop {} {\enquote {\bibinfo {title} {Decomposing
  multivariate information},}\ } (\bibinfo {year} {2010}),\ \Eprint
  {http://arxiv.org/abs/arXiv:1004.2515v1} {arXiv:1004.2515v1} \BibitemShut
  {NoStop}%
\bibitem [{\citenamefont {Cerf}\ and\ \citenamefont {Adami}(1997)}]{Cerf1997}%
  \BibitemOpen
  \bibfield  {author} {\bibinfo {author} {\bibfnamefont {N.~J.}\ \bibnamefont
  {Cerf}}\ and\ \bibinfo {author} {\bibfnamefont {C.}~\bibnamefont {Adami}},\
  }\href@noop {} {\bibfield  {journal} {\bibinfo  {journal} {Phys. Rev. A}\
  }\textbf {\bibinfo {volume} {55}},\ \bibinfo {pages} {3371} (\bibinfo {year}
  {1997})}\BibitemShut {NoStop}%
\bibitem [{\citenamefont {Matsuda}(2000)}]{Matsuda2000}%
  \BibitemOpen
  \bibfield  {author} {\bibinfo {author} {\bibfnamefont {H.}~\bibnamefont
  {Matsuda}},\ }\href@noop {} {\bibfield  {journal} {\bibinfo  {journal} {Phys.
  Rev. E}\ }\textbf {\bibinfo {volume} {62}},\ \bibinfo {pages} {3096}
  (\bibinfo {year} {2000})}\BibitemShut {NoStop}%
\bibitem [{\citenamefont {Anastassiou}(2007)}]{Anastassiou2007}%
  \BibitemOpen
  \bibfield  {author} {\bibinfo {author} {\bibfnamefont {D.}~\bibnamefont
  {Anastassiou}},\ }\href@noop {} {\bibfield  {journal} {\bibinfo  {journal}
  {Molecular Systems Biology}\ }\textbf {\bibinfo {volume} {3}},\ \bibinfo
  {pages} {83} (\bibinfo {year} {2007})}\BibitemShut {NoStop}%
\bibitem [{\citenamefont {Chanda}\ \emph {et~al.}(2007)\citenamefont {Chanda},
  \citenamefont {Zhang}, \citenamefont {Brazeau}, \citenamefont {Sucheston},
  \citenamefont {Freudenheim}, \citenamefont {Ambrosone},\ and\ \citenamefont
  {Ramanathan}}]{Chanda2007}%
  \BibitemOpen
  \bibfield  {author} {\bibinfo {author} {\bibfnamefont {P.}~\bibnamefont
  {Chanda}}, \bibinfo {author} {\bibfnamefont {A.}~\bibnamefont {Zhang}},
  \bibinfo {author} {\bibfnamefont {D.}~\bibnamefont {Brazeau}}, \bibinfo
  {author} {\bibfnamefont {L.}~\bibnamefont {Sucheston}}, \bibinfo {author}
  {\bibfnamefont {J.~L.}\ \bibnamefont {Freudenheim}}, \bibinfo {author}
  {\bibfnamefont {C.}~\bibnamefont {Ambrosone}}, \ and\ \bibinfo {author}
  {\bibfnamefont {M.}~\bibnamefont {Ramanathan}},\ }\href@noop {} {\bibfield
  {journal} {\bibinfo  {journal} {American Journal of Human Genetics}\ }\textbf
  {\bibinfo {volume} {81}},\ \bibinfo {pages} {939} (\bibinfo {year}
  {2007})}\BibitemShut {NoStop}%
\bibitem [{\citenamefont {Brenner}\ \emph {et~al.}(2000)\citenamefont
  {Brenner}, \citenamefont {Strong}, \citenamefont {Koberle}, \citenamefont
  {Bialek},\ and\ \citenamefont {de~Ruyter~van Steveninck}}]{Brenner2000}%
  \BibitemOpen
  \bibfield  {author} {\bibinfo {author} {\bibfnamefont {N.}~\bibnamefont
  {Brenner}}, \bibinfo {author} {\bibfnamefont {S.~P.}\ \bibnamefont {Strong}},
  \bibinfo {author} {\bibfnamefont {R.}~\bibnamefont {Koberle}}, \bibinfo
  {author} {\bibfnamefont {W.}~\bibnamefont {Bialek}}, \ and\ \bibinfo {author}
  {\bibfnamefont {R.~R.}\ \bibnamefont {de~Ruyter~van Steveninck}},\
  }\href@noop {} {\bibfield  {journal} {\bibinfo  {journal} {Neural
  Computation}\ }\textbf {\bibinfo {volume} {12}},\ \bibinfo {pages} {1531}
  (\bibinfo {year} {2000})}\BibitemShut {NoStop}%
\bibitem [{\citenamefont {Schneidman}\ \emph
  {et~al.}(2003{\natexlab{b}})\citenamefont {Schneidman}, \citenamefont
  {Bialek},\ and\ \citenamefont {{Berry II}}}]{SchneidmanBialek2003}%
  \BibitemOpen
  \bibfield  {author} {\bibinfo {author} {\bibfnamefont {E.}~\bibnamefont
  {Schneidman}}, \bibinfo {author} {\bibfnamefont {W.}~\bibnamefont {Bialek}},
  \ and\ \bibinfo {author} {\bibfnamefont {M.~J.}\ \bibnamefont {{Berry II}}},\
  }\href@noop {} {\bibfield  {journal} {\bibinfo  {journal} {Journal of
  Neuroscience}\ }\textbf {\bibinfo {volume} {23}},\ \bibinfo {pages} {11539}
  (\bibinfo {year} {2003}{\natexlab{b}})}\BibitemShut {NoStop}%
\bibitem [{\citenamefont {Bettencourt}\ \emph {et~al.}(2007)\citenamefont
  {Bettencourt}, \citenamefont {Stephens}, \citenamefont {Ham},\ and\
  \citenamefont {Gross}}]{Bettencourt2007}%
  \BibitemOpen
  \bibfield  {author} {\bibinfo {author} {\bibfnamefont {L.~M.~A.}\
  \bibnamefont {Bettencourt}}, \bibinfo {author} {\bibfnamefont {G.~J.}\
  \bibnamefont {Stephens}}, \bibinfo {author} {\bibfnamefont {M.~I.}\
  \bibnamefont {Ham}}, \ and\ \bibinfo {author} {\bibfnamefont {G.~W.}\
  \bibnamefont {Gross}},\ }\href@noop {} {\bibfield  {journal} {\bibinfo
  {journal} {Phys. Rev. E}\ }\textbf {\bibinfo {volume} {75}},\ \bibinfo
  {pages} {021915} (\bibinfo {year} {2007})}\BibitemShut {NoStop}%
\bibitem [{\citenamefont {Tononi}\ \emph {et~al.}(1994)\citenamefont {Tononi},
  \citenamefont {Sporns},\ and\ \citenamefont {Edelman}}]{Tononi1994}%
  \BibitemOpen
  \bibfield  {author} {\bibinfo {author} {\bibfnamefont {G.}~\bibnamefont
  {Tononi}}, \bibinfo {author} {\bibfnamefont {O.}~\bibnamefont {Sporns}}, \
  and\ \bibinfo {author} {\bibfnamefont {G.~M.}\ \bibnamefont {Edelman}},\
  }\href@noop {} {\bibfield  {journal} {\bibinfo  {journal} {Proceedings of the
  National Academy of Sciences}\ }\textbf {\bibinfo {volume} {91}},\ \bibinfo
  {pages} {5033} (\bibinfo {year} {1994})}\BibitemShut {NoStop}%
\bibitem [{\citenamefont {Sporns}\ \emph {et~al.}(2000)\citenamefont {Sporns},
  \citenamefont {Tononi},\ and\ \citenamefont {Edelman}}]{Sporns2000}%
  \BibitemOpen
  \bibfield  {author} {\bibinfo {author} {\bibfnamefont {O.}~\bibnamefont
  {Sporns}}, \bibinfo {author} {\bibfnamefont {G.}~\bibnamefont {Tononi}}, \
  and\ \bibinfo {author} {\bibfnamefont {G.~E.}\ \bibnamefont {Edelman}},\
  }\href@noop {} {\bibfield  {journal} {\bibinfo  {journal} {Cerebral Cortex}\
  }\textbf {\bibinfo {volume} {10}},\ \bibinfo {pages} {127} (\bibinfo {year}
  {2000})}\BibitemShut {NoStop}%
\bibitem [{\citenamefont {Wennekers}\ and\ \citenamefont
  {Ay}(2003)}]{Wennekers2003}%
  \BibitemOpen
  \bibfield  {author} {\bibinfo {author} {\bibfnamefont {T.}~\bibnamefont
  {Wennekers}}\ and\ \bibinfo {author} {\bibfnamefont {N.}~\bibnamefont {Ay}},\
  }\href@noop {} {\bibfield  {journal} {\bibinfo  {journal} {Theory in
  Bioscience}\ }\textbf {\bibinfo {volume} {122}},\ \bibinfo {pages} {5}
  (\bibinfo {year} {2003})}\BibitemShut {NoStop}%
\bibitem [{\citenamefont {Timme}()}]{TimmeWebsite}%
  \BibitemOpen
  \bibfield  {author} {\bibinfo {author} {\bibfnamefont {N.}~\bibnamefont
  {Timme}},\ }\href@noop {} {}\bibinfo {note}
  {Http://mypage.iu.edu/$\sim$nmtimme}\BibitemShut {NoStop}%
\bibitem [{\citenamefont {Timme}\ \emph {et~al.}(2011)\citenamefont {Timme},
  \citenamefont {Alford}, \citenamefont {Flecker},\ and\ \citenamefont
  {Beggs}}]{Timme2011}%
  \BibitemOpen
  \bibfield  {author} {\bibinfo {author} {\bibfnamefont {N.}~\bibnamefont
  {Timme}}, \bibinfo {author} {\bibfnamefont {W.}~\bibnamefont {Alford}},
  \bibinfo {author} {\bibfnamefont {B.}~\bibnamefont {Flecker}}, \ and\
  \bibinfo {author} {\bibfnamefont {J.~M.}\ \bibnamefont {Beggs}},\ }\href@noop
  {} {\enquote {\bibinfo {title} {Multivariate information measures: an
  experimentalist's perspective},}\ } (\bibinfo {year} {2011}),\ \Eprint
  {http://arxiv.org/abs/arXiv:1111.6857v4} {arXiv:1111.6857v4} \BibitemShut
  {NoStop}%
\bibitem [{\citenamefont {Griffith}\ and\ \citenamefont
  {Koch}(2011)}]{Griffith2011}%
  \BibitemOpen
  \bibfield  {author} {\bibinfo {author} {\bibfnamefont {V.}~\bibnamefont
  {Griffith}}\ and\ \bibinfo {author} {\bibfnamefont {C.}~\bibnamefont
  {Koch}},\ }\href@noop {} {\enquote {\bibinfo {title} {Quantifying synergistic
  mutual information},}\ } (\bibinfo {year} {2011}),\ \Eprint
  {http://arxiv.org/abs/arXiv:1112.1680v5} {arXiv:1112.1680v5} \BibitemShut
  {NoStop}%
\bibitem [{\citenamefont {Cover}\ and\ \citenamefont
  {Thomas}(2006)}]{Cover2006}%
  \BibitemOpen
  \bibfield  {author} {\bibinfo {author} {\bibfnamefont {T.~M.}\ \bibnamefont
  {Cover}}\ and\ \bibinfo {author} {\bibfnamefont {J.~A.}\ \bibnamefont
  {Thomas}},\ }\href@noop {} {\emph {\bibinfo {title} {Elements of information
  theory}}},\ \bibinfo {edition} {2nd}\ ed.\ (\bibinfo  {publisher}
  {Wiley-Interscience},\ \bibinfo {year} {2006})\BibitemShut {NoStop}%
\bibitem [{Note1()}]{Note1}%
  \BibitemOpen
  \bibinfo {note} {Throughout the paper we will use capital letters to refer to
  variables and lower case letters to refer to individual values of those
  variables. We will also use discrete variables, though several of the
  information measures discussed can be directly extended to continuous
  variables. When working with a continuous variable, various techniques
  exists, such as kernel density estimation, which can be used to infer a
  discrete distribution from a continuous variable. Logarithms will be base 2
  throughout in order to produce information values in units of
  bits.}\BibitemShut {Stop}%
\bibitem [{Note2()}]{Note2}%
  \BibitemOpen
  \bibinfo {note} {We will use S to refer to a set of $N$ $X$ variables such
  that $S = \protect \{X_1, X_2, \protect \dots X_N\protect \}$ throughout the
  paper.}\BibitemShut {Stop}%
\bibitem [{\citenamefont {Bettencourt}\ \emph {et~al.}(2008)\citenamefont
  {Bettencourt}, \citenamefont {Gintautas},\ and\ \citenamefont
  {Ham}}]{Bettencourt2008}%
  \BibitemOpen
  \bibfield  {author} {\bibinfo {author} {\bibfnamefont {L.~M.~A.}\
  \bibnamefont {Bettencourt}}, \bibinfo {author} {\bibfnamefont
  {V.}~\bibnamefont {Gintautas}}, \ and\ \bibinfo {author} {\bibfnamefont
  {M.~I.}\ \bibnamefont {Ham}},\ }\href@noop {} {\bibfield  {journal} {\bibinfo
   {journal} {Physical Review Letters}\ }\textbf {\bibinfo {volume} {100}},\
  \bibinfo {pages} {238701} (\bibinfo {year} {2008})}\BibitemShut {NoStop}%
\bibitem [{\citenamefont {Gat}\ and\ \citenamefont {Tishby}()}]{Gat1999}%
  \BibitemOpen
  \bibfield  {author} {\bibinfo {author} {\bibfnamefont {I.}~\bibnamefont
  {Gat}}\ and\ \bibinfo {author} {\bibfnamefont {N.}~\bibnamefont {Tishby}},\
  }in\ \href@noop {} {\emph {\bibinfo {booktitle} {Neural Information
  Processing Systems 11}}},\ \bibinfo {editor} {edited by\ \bibinfo {editor}
  {\bibfnamefont {M.~S.}\ \bibnamefont {Kearns}}, \bibinfo {editor}
  {\bibfnamefont {S.~A.}\ \bibnamefont {Solla}}, \ and\ \bibinfo {editor}
  {\bibfnamefont {D.~A.}\ \bibnamefont {Cohn}}}\ (\bibinfo  {publisher} {MIT
  Press})\ p.\ \bibinfo {pages} {111}\BibitemShut {NoStop}%
\bibitem [{\citenamefont {Jakulin}\ and\ \citenamefont
  {Bratko}(2008)}]{Jakulin2008}%
  \BibitemOpen
  \bibfield  {author} {\bibinfo {author} {\bibfnamefont {A.}~\bibnamefont
  {Jakulin}}\ and\ \bibinfo {author} {\bibfnamefont {I.}~\bibnamefont
  {Bratko}},\ }\href@noop {} {\enquote {\bibinfo {title} {Quantifying and
  visualizing attribute interactions},}\ } (\bibinfo {year} {2008}),\ \Eprint
  {http://arxiv.org/abs/arXiv:cs/0308002v3} {arXiv:cs/0308002v3} \BibitemShut
  {NoStop}%
\bibitem [{\citenamefont {Bell}(2003)}]{Bell2003}%
  \BibitemOpen
  \bibfield  {author} {\bibinfo {author} {\bibfnamefont {A.~J.}\ \bibnamefont
  {Bell}},\ }in\ \href@noop {} {\emph {\bibinfo {booktitle} {International
  workshop on independent component analysis and blind signal separation}}}\
  (\bibinfo {year} {2003})\ p.\ \bibinfo {pages} {921}\BibitemShut {NoStop}%
\bibitem [{\citenamefont {Han}(1978)}]{Han1978}%
  \BibitemOpen
  \bibfield  {author} {\bibinfo {author} {\bibfnamefont {T.~S.}\ \bibnamefont
  {Han}},\ }\href@noop {} {\bibfield  {journal} {\bibinfo  {journal}
  {Information and Control}\ }\textbf {\bibinfo {volume} {36}},\ \bibinfo
  {pages} {133} (\bibinfo {year} {1978})}\BibitemShut {NoStop}%
\bibitem [{\citenamefont {Abdallah}\ and\ \citenamefont
  {Plumbley}(2010)}]{Abdallah2010}%
  \BibitemOpen
  \bibfield  {author} {\bibinfo {author} {\bibfnamefont {S.~A.}\ \bibnamefont
  {Abdallah}}\ and\ \bibinfo {author} {\bibfnamefont {M.~D.}\ \bibnamefont
  {Plumbley}},\ }\href@noop {} {\enquote {\bibinfo {title} {A measure of
  statistical complexity based on predictive information},}\ } (\bibinfo {year}
  {2010}),\ \Eprint {http://arxiv.org/abs/arXiv:1012.1890v1}
  {arXiv:1012.1890v1} \BibitemShut {NoStop}%
\bibitem [{\citenamefont {Olbrich}\ \emph {et~al.}(2008)\citenamefont
  {Olbrich}, \citenamefont {Bertschinger}, \citenamefont {Ay},\ and\
  \citenamefont {Jost}}]{Olbrich2008}%
  \BibitemOpen
  \bibfield  {author} {\bibinfo {author} {\bibfnamefont {E.}~\bibnamefont
  {Olbrich}}, \bibinfo {author} {\bibfnamefont {N.}~\bibnamefont
  {Bertschinger}}, \bibinfo {author} {\bibfnamefont {N.}~\bibnamefont {Ay}}, \
  and\ \bibinfo {author} {\bibfnamefont {J.}~\bibnamefont {Jost}},\ }\href@noop
  {} {\bibfield  {journal} {\bibinfo  {journal} {European Physical Journal B}\
  }\textbf {\bibinfo {volume} {63}},\ \bibinfo {pages} {407} (\bibinfo {year}
  {2008})}\BibitemShut {NoStop}%
\bibitem [{\citenamefont {Latham}\ and\ \citenamefont
  {Nirenberg}(2005)}]{Latham2005}%
  \BibitemOpen
  \bibfield  {author} {\bibinfo {author} {\bibfnamefont {P.~E.}\ \bibnamefont
  {Latham}}\ and\ \bibinfo {author} {\bibfnamefont {S.}~\bibnamefont
  {Nirenberg}},\ }\href@noop {} {\bibfield  {journal} {\bibinfo  {journal}
  {Journal of Neuroscience}\ }\textbf {\bibinfo {volume} {25}},\ \bibinfo
  {pages} {5195} (\bibinfo {year} {2005})}\BibitemShut {NoStop}%
\bibitem [{\citenamefont {James}\ \emph {et~al.}(2011)\citenamefont {James},
  \citenamefont {Ellison},\ and\ \citenamefont {Crutchfield}}]{James2011}%
  \BibitemOpen
  \bibfield  {author} {\bibinfo {author} {\bibfnamefont {R.~G.}\ \bibnamefont
  {James}}, \bibinfo {author} {\bibfnamefont {C.~J.}\ \bibnamefont {Ellison}},
  \ and\ \bibinfo {author} {\bibfnamefont {J.~P.}\ \bibnamefont
  {Crutchfield}},\ }\href@noop {} {\bibfield  {journal} {\bibinfo  {journal}
  {Chaos}\ }\textbf {\bibinfo {volume} {21}},\ \bibinfo {pages} {037109}
  (\bibinfo {year} {2011})}\BibitemShut {NoStop}%
\bibitem [{\citenamefont {Flecker}\ \emph {et~al.}(2011)\citenamefont
  {Flecker}, \citenamefont {Alford}, \citenamefont {Beggs}, \citenamefont
  {Williams},\ and\ \citenamefont {Beer}}]{Flecker2011}%
  \BibitemOpen
  \bibfield  {author} {\bibinfo {author} {\bibfnamefont {B.}~\bibnamefont
  {Flecker}}, \bibinfo {author} {\bibfnamefont {W.}~\bibnamefont {Alford}},
  \bibinfo {author} {\bibfnamefont {J.~M.}\ \bibnamefont {Beggs}}, \bibinfo
  {author} {\bibfnamefont {P.~L.}\ \bibnamefont {Williams}}, \ and\ \bibinfo
  {author} {\bibfnamefont {R.~D.}\ \bibnamefont {Beer}},\ }\href@noop {}
  {\bibfield  {journal} {\bibinfo  {journal} {Chaos}\ }\textbf {\bibinfo
  {volume} {21}},\ \bibinfo {pages} {037104} (\bibinfo {year}
  {2011})}\BibitemShut {NoStop}%
\bibitem [{\citenamefont {DeWeese}\ and\ \citenamefont
  {Meister}(1999)}]{DeWeese1999}%
  \BibitemOpen
  \bibfield  {author} {\bibinfo {author} {\bibfnamefont {M.~R.}\ \bibnamefont
  {DeWeese}}\ and\ \bibinfo {author} {\bibfnamefont {M.}~\bibnamefont
  {Meister}},\ }\href@noop {} {\bibfield  {journal} {\bibinfo  {journal}
  {Network: Computation in Neural Systems}\ }\textbf {\bibinfo {volume} {10}},\
  \bibinfo {pages} {325} (\bibinfo {year} {1999})}\BibitemShut {NoStop}%
\bibitem [{Note3()}]{Note3}%
  \BibitemOpen
  \bibinfo {note} {It should be noted that DeWeese and Meister refer to the
  expression in Eq. (\ref {EQ25}) as the specific surprise.}\BibitemShut
  {Stop}%
\bibitem [{\citenamefont {Wagenaar}\ \emph
  {et~al.}(2006{\natexlab{a}})\citenamefont {Wagenaar}, \citenamefont {Pine},\
  and\ \citenamefont {Potter}}]{WagenaarPine2006}%
  \BibitemOpen
  \bibfield  {author} {\bibinfo {author} {\bibfnamefont {D.~A.}\ \bibnamefont
  {Wagenaar}}, \bibinfo {author} {\bibfnamefont {J.}~\bibnamefont {Pine}}, \
  and\ \bibinfo {author} {\bibfnamefont {S.~M.}\ \bibnamefont {Potter}},\
  }\href@noop {} {\bibfield  {journal} {\bibinfo  {journal} {BMC Neuroscience}\
  }\textbf {\bibinfo {volume} {7}} (\bibinfo {year}
  {2006}{\natexlab{a}})}\BibitemShut {NoStop}%
\bibitem [{\citenamefont {Kamioka}\ \emph {et~al.}(1996)\citenamefont
  {Kamioka}, \citenamefont {Maeda}, \citenamefont {Jimbo}, \citenamefont
  {Robinson},\ and\ \citenamefont {Kawana}}]{Kamioka1996}%
  \BibitemOpen
  \bibfield  {author} {\bibinfo {author} {\bibfnamefont {H.}~\bibnamefont
  {Kamioka}}, \bibinfo {author} {\bibfnamefont {E.}~\bibnamefont {Maeda}},
  \bibinfo {author} {\bibfnamefont {Y.}~\bibnamefont {Jimbo}}, \bibinfo
  {author} {\bibfnamefont {H.~P.~C.}\ \bibnamefont {Robinson}}, \ and\ \bibinfo
  {author} {\bibfnamefont {A.}~\bibnamefont {Kawana}},\ }\href@noop {}
  {\bibfield  {journal} {\bibinfo  {journal} {Neuroscience Letters}\ }\textbf
  {\bibinfo {volume} {206}},\ \bibinfo {pages} {109} (\bibinfo {year}
  {1996})}\BibitemShut {NoStop}%
\bibitem [{\citenamefont {Wagenaar}\ \emph
  {et~al.}(2006{\natexlab{b}})\citenamefont {Wagenaar}, \citenamefont
  {Nadasdy},\ and\ \citenamefont {Potter}}]{WagenaarNadasdy2006}%
  \BibitemOpen
  \bibfield  {author} {\bibinfo {author} {\bibfnamefont {D.~A.}\ \bibnamefont
  {Wagenaar}}, \bibinfo {author} {\bibfnamefont {Z.}~\bibnamefont {Nadasdy}}, \
  and\ \bibinfo {author} {\bibfnamefont {S.~M.}\ \bibnamefont {Potter}},\
  }\href@noop {} {\bibfield  {journal} {\bibinfo  {journal} {Physical Review
  E}\ }\textbf {\bibinfo {volume} {73}},\ \bibinfo {pages} {051907} (\bibinfo
  {year} {2006}{\natexlab{b}})}\BibitemShut {NoStop}%
\bibitem [{\citenamefont {Pasquale}\ \emph {et~al.}(2008)\citenamefont
  {Pasquale}, \citenamefont {Massobrio}, \citenamefont {Bologna}, \citenamefont
  {Chiappalonea},\ and\ \citenamefont {Martinoia}}]{Pasquale2008}%
  \BibitemOpen
  \bibfield  {author} {\bibinfo {author} {\bibfnamefont {V.}~\bibnamefont
  {Pasquale}}, \bibinfo {author} {\bibfnamefont {P.}~\bibnamefont {Massobrio}},
  \bibinfo {author} {\bibfnamefont {L.~L.}\ \bibnamefont {Bologna}}, \bibinfo
  {author} {\bibfnamefont {M.}~\bibnamefont {Chiappalonea}}, \ and\ \bibinfo
  {author} {\bibfnamefont {S.}~\bibnamefont {Martinoia}},\ }\href@noop {}
  {\bibfield  {journal} {\bibinfo  {journal} {Neuroscience}\ }\textbf {\bibinfo
  {volume} {153}},\ \bibinfo {pages} {1354} (\bibinfo {year}
  {2008})}\BibitemShut {NoStop}%
\bibitem [{\citenamefont {Tetzlaff}\ \emph {et~al.}(2010)\citenamefont
  {Tetzlaff}, \citenamefont {Okujeni}, \citenamefont {Egert}, \citenamefont
  {Worgotter},\ and\ \citenamefont {Butz}}]{Tetzlaff2010}%
  \BibitemOpen
  \bibfield  {author} {\bibinfo {author} {\bibfnamefont {C.}~\bibnamefont
  {Tetzlaff}}, \bibinfo {author} {\bibfnamefont {S.}~\bibnamefont {Okujeni}},
  \bibinfo {author} {\bibfnamefont {U.}~\bibnamefont {Egert}}, \bibinfo
  {author} {\bibfnamefont {F.}~\bibnamefont {Worgotter}}, \ and\ \bibinfo
  {author} {\bibfnamefont {M.}~\bibnamefont {Butz}},\ }\href@noop {} {\bibfield
   {journal} {\bibinfo  {journal} {PLoS Computationa Biology}\ }\textbf
  {\bibinfo {volume} {6}},\ \bibinfo {pages} {e1001013} (\bibinfo {year}
  {2010})}\BibitemShut {NoStop}%
\end{thebibliography}%

\end{document}